\documentclass[preprint,amsmath,amssymb,showpacs]{revtex4}
\usepackage{graphicx}
\newcommand{\ks} {{\bf k}}
\newcommand{\ac} {{\bf A}}
\newcommand{\as} {{\bf a}}
\newcommand{\bs} {{\bf b}}
\newcommand{\jc} {{\bf J}}
\newcommand{\ec} {{\bf E}}
\newcommand{\es} {{\bf e}}
\newcommand{\pc} {{\bf P}}
\newcommand{\gc} {{\bf G}}
\newcommand{\rc} {{\bf R}}
\newcommand{\rs} {{\bf r}}
\newcommand{\hs} {{\bf h}}

\newcommand{\eps} {\epsilon}

\newcommand{\um}[4]{\left(\begin{array}{cc}
#1 & #2  \\
#3 & #4 
\end{array}\right)}

\begin{document}
\title{High-harmonic generation in zinc oxide subjected to intense mid-infrared femtosecond laser pulse}
\author{Boyan Obreshkov$^1$ and Tzveta Apostolova$^{1,2}$}
\affiliation{$^1$ Institute for Nuclear Research and Nuclear Energy,
Bulgarian Academy of Sciences, Tsarigradsko chausse\'{e} 72, Sofia
1784, Bulgaria}
\affiliation{$^2$ Institute for Advanced Physical Studies, New Bulgarian University 
, Montevideo 21,  Sofia 1618, Bulgaria}
\begin{abstract}

We theoretically investigate photo-excitation of electron-hole pairs and high harmonic generation in the bulk of zinc oxide (ZnO) subjected to intense femto-second laser pulses with mid-infrared wavelength. The main microscopic mechanism of solid-state HHG is identified by separating resonant from non-resonant non-linear optical responses in the photo-excited solid. It allows us
to obtain an effective description of the light-matter interaction 
in which electrons become subject to weak atto-second pulse train with the second harmonic of the drive laser frequency being the repetition frequency in the train. Under a condition of constructive interference between electronic transitions at each half-cycle of the drive laser pulse, resonant-like excitation of electron-hole pairs occurs, analogously to above threshold ionization in a gas phase. The inter-band motion of charge carriers creates rapidly oscillating electric dipole moment, which emits radiation in the form of high-order harmonics of the drive laser frequency. We also discuss the importance of the pulse envelope for producing clean frequency combs as observed in experiments. Good semi-quantitative agreement with the experimental data is found: clean and well defined odd-order harmonic peaks extending well beyond the band edge of ZnO  are exhibited for laser linearly polarized at right angles to the optical axis of the crystal. 
\end{abstract}

\maketitle

\section{Introduction} 

High-order harmonic generation (HHG) in a bulk crystal was observed for the first time after the irradiation of zinc-oxide with intense mid-infrared (MIR) few-cycle laser pulse \cite{Ghimire2011}. High order harmonics up to the 25th order were produced in the transmission through a 500 $\mu$m thick ZnO crystal at peak laser intensity ~5 $\times 10^{12}$ W/cm$^2$. The measured harmonic spectra extended beyond the band edge of ZnO and displayed distinct odd harmonic peaks superimposed on a continuous  background. The harmonics spectra display non-perturbative signatures:  i) the cutoff energy for solid state HHG scales linearly with the peak electric field of the pulsed laser, which is in contrast to the gas-phase harmonics, where the cut-off scales linearly with the peak laser intensity and ii)  the dependence of the spectral intensity of individual harmonics on the peak laser intensity of the driving pulse was found to substantially deviate from a power law expected from perturbative nonlinear optics.  It was also found that high-order harmonic spectra depend very sensitively on the laser polarization direction relative to the crystallographic axes. To explain their results \cite{Ghimire2011} the authors proposed a one-band model in which harmonics of the fundamental laser frequency are a result of laser-driven Bloch oscillations of conduction electrons. Thus the non-linear band dispersion of the group velocity of electrons was assumed to be the main mechanism for solid state HHG. 

Subsequently solid-state HHG were experimentally achieved in other wide bandgap dielectrics, in silica  \cite{Luu2015} and in gallium selenide \cite{Schubert2014,Hohenleutner2015}. In these  experiments, the intra-band mechanism of HHG by dynamical Bloch oscillations of conduction electrons was verified. 

While HHG experiments in solids display clean and well-defined harmonic spectra with distinct peaks at multiples of the driving laser frequency, the predictions of the corresponding theoretical models for solid-state HHG often lack clear harmonic spectra.  For instance, in the framework of two band model for laser-driven ZnO \cite{Vampa2014,Vampa2015} (analogous to the semi-classical three-step model of HHG in the gas phase \cite{Lewenstein1994,Schafer2015,Corkum1990}), an inter-band mechanism of build-up of transient polarization in the photo-excited solid was identified as a primary source of HHG, which contrasts the conclusions in \cite{Ghimire2011}. However, in order to reproduce the clean frequency comb observed in the ZnO experiment, one has to phenomenologically introduce an ultra-fast polarization de-phasing process. But a physical mechanism associated with such ultrashort de-phasing time-scale, which is fraction of an optical cycle, has not been clearly identified  \cite{Floss2018,Orlando2020,Freeman2022,Brown2024}.  

Theoretical analysis of HHG in bulk solids is often made in the framework of two-band models \cite{Vampa2014,Tamaya2016,Yue2020}, while the spectral characteristics of HHG is usually understood through the semi-classical three-step model in terms of semi-classical trajectories of electron-hole pairs. Solid state HHG involving multiple bands was also discussed in, e.g. \cite{Hawkins2015,Wu2015,Guan2016,Dejean2017,Navarette2019}. Since peculiarities of the static band structure affect spectral and temporal properties of HHG, the mechanism for HHG in solids should be analyzed within a realistic model for the crystal band structure.

Though much theoretical effort has been devoted to understanding the spectral properties of solid-state HHG within the semi-classical three-step model, the corresponding analyses were less focused onto the time-dependent characteristics of the photo-current of electron and hole pairs.  Much earlier, Brunel considered HHG in a gas undergoing tunnel ionization by strong and low-frequency laser fields \cite{Brunel1990}. When photo-ionization takes place, free electrons emerge in the continuum in short temporal bursts around the peaks of the laser electric field. The sharp transient features of the electron density distribution translate into the temporal profile of the photo-induced electric current, which was found to be responsible for the emission of odd-order harmonics in a gas phase.  

In this paper we theoretically investigate photo-excitation of electron-hole pairs and high-harmonic generation in ZnO subjected to intense few-cycle femto-second laser pulses with wavelength in the range 2-8 $\mu$m, and the peak laser intensity was varied in the range $10^{11}-10^{14}$ W/cm$^2$. The laser polarization direction was either parallel or perpendicular to the optical axis of the ZnO crystal.  In velocity gauge, the time-dependent Sch\"{o}dinger equation in single active electron approximation was employed to investigate HHG in the bulk of ZnO.  The paper is organized as follows:  Sec.II includes description of our theoretical model, Sec. III presents results and discussion of electron-hole pair excitation, the time evolution of the conduction electron density and the transient photo-currents associated to solid-state HHG, and Sec.IV includes our main conclusions. Unless otherwise stated, atomic units are used throughout this paper ($e=\hbar=m_e=1$). 

\section{Theoretical model}

\subsection{Ground state electronic structure} 

The ground state electronic structure of zinc oxide is described by the empirical pseudo-potential method \cite{Cohen1966}. The wurtzite structure of ZnO exhibits a hexagonal lattice with primitive lattice translation vectors 
\begin{equation}
 \as_1=(0.5 a,-\sqrt{3}a/2,0), \quad \as_2=(0.5a,\sqrt{3}a/2,0), \quad 
 \as_3=(0,0,c) 
\end{equation}
where $a$ and $c$ are the two lattice constants. The position vectors of the two zinc atoms in the crystal unit cell are $\rc_1=\as_1/3 + 2 \as_2/3$ and $\rc_2=2 \as_1/3 + \as_2/3+\as_3/2$, and the position vectors of the two oxygen atoms are obtained from them by translation along the optical axis $\rc_3=\rc_1+ u \as_3$ and $\rc_4=\rc_2+ u \as_3$, here the  displacement parameter $u=(a/c)^2$ (in ideal wurtzite structure). 
The primitive reciprocal lattice wave-vectors are  
\begin{equation}
 \bs_1=\frac{2 \pi}{a} (1,-1/\sqrt{3},0), \quad \bs_2=\frac{2 \pi}{a} (1,1/\sqrt{3},0), \quad 
 \bs_3= \frac{2 \pi}{c} (0,0,1) 
\end{equation}
The screened electron-ion pseudo-potential for wurtzite is expanded in plane waves in the basis of the reciprocal lattice wave vectors ($\gc=m_1 \bs_1 + m_2 \bs_2 + m_3 \bs_3$)  
\begin{equation}
 V(\rs)= \sum_{\gc} [S_S(\gc) V_S(G) + i S_A(\gc) V_A(G)]e^{i \gc \cdot \rs}
\end{equation}
The symmetric and anti-symmetric form factors for ZnO are 
\begin{equation}
V_S(G)= [V_{Zn}(G)+V_O(G)]/2, \quad  V_A(G)= [V_{Zn}(G)+V_O(G)]/2
\end{equation}
and the symmetric and anti-symmetric structure factors can be written as 
\begin{eqnarray*}
& & S_S(\gc)= \cos( \gc \cdot (\rc_1-\rc_2)/2) \cos(u c G_3/2)
\nonumber \\
& & S_A(\gc)= \cos( \gc \cdot (\rc_1-\rc_2)/2) 
\sin(u c G_3/2)
\end{eqnarray*}
The form-factors for zinc and oxygen atoms were presented by a smooth function of the form \cite{Schluter1975} 
\begin{equation}
 V(q) = \frac{v_1 (q^2-v_2)}{\exp(v_3(q^2-v_4))+1} 
\end{equation}
in terms of eight pseudo-potential parameters, which according to \cite{Fan2006} are: $v_1=0.11003,v_2=1.79208,v_3=0.7708,v_4=4.25370$ for a zinc atom, and $v_1=0.19629,v_2=4.90951,v_3=1.24475,v_4=3.60095$ for an oxygen atom; all pseudo-potential parameters are given in Rydberg units. The lattice constants $a$ and $c$ used in our calculation are $3.25$ and $5.21$ \AA,~respectively, and the $u$ parameter was set to $u=0.37$. The static band structure of ZnO is found by solving for the eigen-energies $\varepsilon_{n \ks}$ and eigen-states $|n \ks \rangle=\sum_{\gc} |\gc+\ks \rangle C_{n,\gc+\ks}$ of the Hamiltonian of a Bloch electron with crystal momentum $\ks$ in a plane-wave basis with high-energy cutoff $E_{cut}=25$ Ry : 
\begin{equation}
 \sum_{\gc'} H_{\gc+\ks,\gc'+\ks} C_{n, \gc'+\ks} =\varepsilon_{n \ks} C_{n, \gc+\ks}
\end{equation}
with the Hamiltonian 
\begin{equation}
 H_{\gc+\ks,\gc'+\ks}= \frac{1}{2} (\gc+\ks)^2 \delta_{\gc, \gc'} + V(\gc-\gc'),
\end{equation}
here the crystal momentum $\ks$ extends over the first Brillouin zone. The Brillouin zone was sampled by a Monkhorst-Pack scheme, according to 
\begin{equation}
 \ks= u_1 \bs_1 + u_2 \bs_2 + u_3 \bs_3 
\end{equation}
where $u_i=(2n_i-N_i-1)/2N_i$ for $n_i=1,\ldots,N_i$ and $i=1,2,3$.

\subsection{Theoretical formalism} 

The time-dependent Hamiltonian of a Bloch electron interacting with spatially uniform laser electric field $\ec(t)$ is obtained from the static one in terms of the kinetic quasi-momentum 
$\ks(t)=\ks+\ac(t)$ 
\begin{equation}
 H(\ks(t))= \sum_n | n \ks(t) \rangle \varepsilon_{n \ks(t)} \langle n \ks(t)| 
 \label{Hkt}
\end{equation}
where $|n \ks(t) \rangle$ are the instantaneous Bloch eigen-states with eigen-energies $\varepsilon_{n \ks(t)}$ and $\ac(t)$ is the electromagnetic  vector potential, related to the laser electric field by $\ec=-\partial_t \ac$. The electrons are assumed to respond as non-interacting particles to the total time-dependent electric field inside the bulk, which is a superposition of the externally applied laser electric field $\ec_{{\rm ext}}$ and the induced electric field $\ec_{{\rm ind}}$ which is a result of the transient polarization of the solid, i.e. $\ec=\ec_{{\rm ext}} +\ec_{{\rm ind}}$, (cf. Refs.\cite{Otobe2008,Apostolova2020}). The external laser electric field is parametrized by 
\begin{equation}
\ec_{{\rm ext}}(t) =\es F \cos(\omega_L t) \sin^2(\pi t/\tau) , \quad 0 \le t \le \tau  
\end{equation}
where $\es$ is the laser polarization unit vector, the peak field strength inside vacuum $F$ is related to the peak laser intensity by $I=F^2$, $\omega_L$ is the laser oscillation frequency corresponding to photon energy $\hbar \omega \approx 0.4$ eV and $\tau$ designates the pulse duration. To obtain the observables associated with HHG, we numerically solve the  single-particle Schr\"{o}dinger equation  
\begin{equation}
 i \partial_t | u_v (\ks,t) \rangle = H(\ks(t)) | u_v (\ks,t) \rangle , \label{Sch}
\end{equation}
subject to the initial conditions $|u_v(\ks,0)\rangle=|v \ks\rangle$, where $v=1 \ldots 8$ and $\ks$ label the eight initially occupied valence electron Bloch states with definite crystal momentum $\ks$; the single-particle density-matrix is defined  
\begin{equation}
 \rho_{\ks}(t) = \sum_{v=1}^8 |u_v (\ks,t) \rangle \langle u_v(\ks,t)|.  
\end{equation}
and the three Cartesian components ($\alpha=x,y,z$) of the transient photo-induced current are obtained from the Brillouin zone integral 
\begin{equation}
 J_{\alpha}(t) =-{{\rm Tr}} \left( \rho(t) h_{\alpha}(t)   \right) =  - \int_{{\rm BZ}} \frac{d^3 \ks}{4 \pi^3} \sum_{nn'} \rho_{nn'\ks}(t) h_{\alpha}(t)_{n' n \ks}
\label{Jkt} 
\end{equation}
here $h_{\alpha}(t)=\partial H(t)/ \partial k_{\alpha} $ is the
field-dependent velocity operator. The current density $\jc$ is related to the transient polarization of the solid by $\jc=d \pc/dt$ and to the induced electric field by $\ec_{{\rm ind}}=-4 \pi \pc$. Eqs.(\ref{Sch}) and (\ref{Jkt}) should be solved self-consistently in a way which is technically equivalent to the time-dependent current density functional theory approach to the response of bulk solids \cite{Maitra2003}. The Fourier transform of the photo-induced current projected onto the laser polarization direction is related to the power spectrum of the emitted radiation inside the bulk of ZnO, which is often associated to high harmonic generation 
\begin{equation}
 S(\omega) = \left|\int dt e^{i \omega t}  \jc(t) \cdot \es \right|^2 
\end{equation}
Details about the numerical method and its implementation are given in appendix.

\section{Numerical results and discussion} 

\subsection{Static band structure and dielectric response}

\begin{figure}
\begin{center}
\includegraphics[width=.5\textwidth]{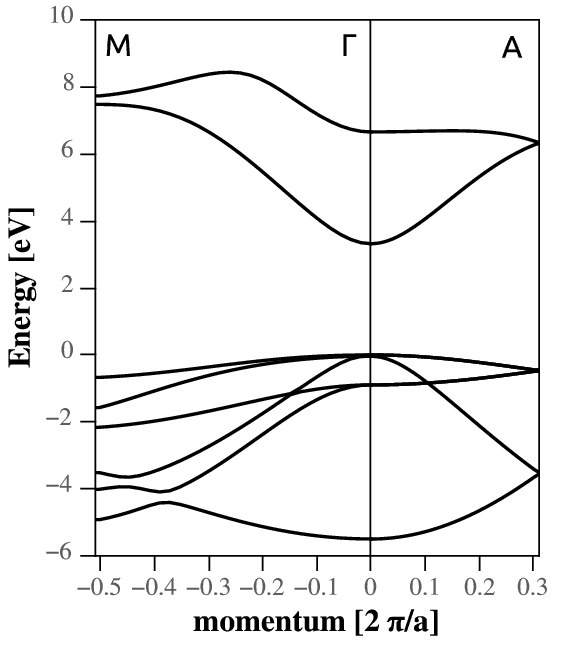}
\caption{Static band structure of zinc oxide 
along the $\Delta$ ($\Gamma$-A) and $\Sigma$ ($\Gamma$-M) 
lines in the Brillouin zone. The crystal momentum is measured in units $2\pi/a$ , and the lattice parameter is $a = 3.25$ \AA.}   
\label{fig:fig1}
\end{center}
\end{figure}

The static band structure of ZnO is shown in Fig.\ref{fig:fig1} along the $\Gamma$-A and $\Gamma$-M lines in the Brillouin zone.
The pseudo-potential method reproduces quantitatively the principal energy gaps and optical properties of ZnO in the UV region. For the MIR wavelength, the direct bandgap energy at the Brillouin center $\Delta=3.34$ eV would require  simultaneous absorption of at least nine photons to produce an electron-hole pair.  The conduction bands of ZnO are mainly derived from the $4s$ orbitals of zinc atoms, while the group of six valence bands stems from the $2p$ orbitals of the oxygen atoms. Around 20 eV below the valence band top, there are two more bands derived from the $2s$ orbitals of the oxygen atoms (not shown). Near the 
$\Gamma$ point, the two degenerate valence bands at the top are derived from $2p_x$ and $2p_y$ orbitals of oxygen atoms, the crystal field split-off $p_z$-like band is slightly below the top of the valence band by $0.04$ eV. Besides the three valence bands at the top, there are three more lower-lying in energy valence bands. The upper two of these lower-lying bands are degenerate at the $\Gamma$ point and are separated by nearly $0.8$ eV from the top. The lowest in energy valence band at the bottom is split by $5.5$ eV from the top. Degeneracy among valence bands is lifted along the $\Gamma$-M line, when all six valence bands are clearly revealed. In this direction in the Brillouin zone, the upper five bands disperse downwards, while the bottom band disperses upwards, such that the group of six bands move close in energy and are strongly coupled. 
 
\begin{figure}
\begin{center}
\includegraphics[width=.5\textwidth]{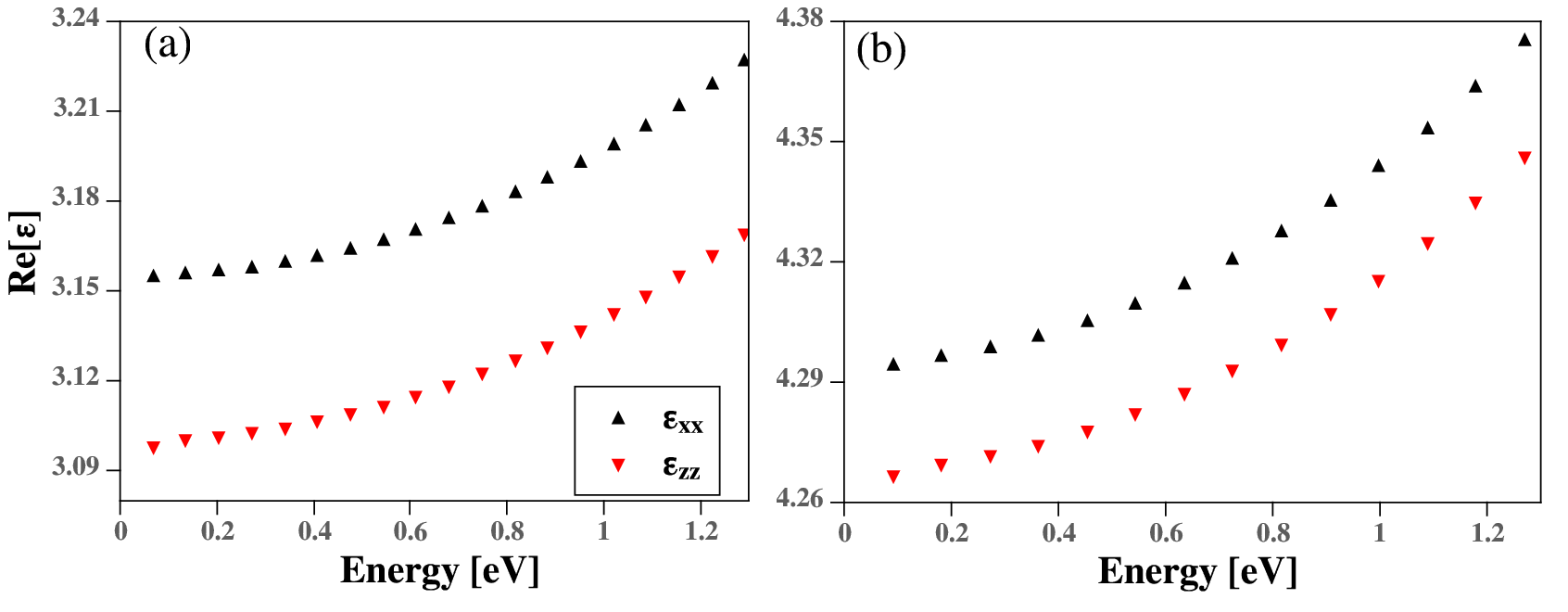}
\caption{Frequency-dependence of the Cartesian components of the  dielectric tensor of ZnO  parallel and perpendicular to the optical axis of the ZnO crystal. In Fig.(a) transitions into and out of the two lowest in energy conduction bands are included, the result in Fig.(b) includes transitions into and out of the eight lowest in energy conduction bands.}   
\label{fig:fig2}
\end{center}
\end{figure}

To test the static band structure and couplings associated with transitions among valence and conduction bands, we verify that the oscillator strength sum-rule is satisfied. Given the Hamiltonian in Eq.(\ref{Hkt}), the Cartesian components of the optical conductivity tensor are evaluated in first order of perturbation theory with the standard expression in velocity gauge  \cite{Passos2018}  
\begin{equation}
\sigma_{\beta \alpha}(\omega)=\frac{i}{\omega} 
\int_{{\rm BZ}} \frac{d^3 \ks}{4 \pi^3} \sum_{nn'} \left( 
\frac{ (h_{\beta})_{n'n \ks} [h_{\alpha},\rho_0]_{nn'\ks}}
{\omega-\omega_{nn'\ks}} + (h_{\alpha \beta})_{n'n \ks} (\rho_0)_{nn'\ks} \right)  
\label{sigma} 
\end{equation}
where in the $\ks$-representation, the ground-state density matrix $\rho_0$ is diagonal with elements $f_{n\ks} \in \{0,1\}$ - the occupation numbers of unperturbed Bloch states, the transition frequencies are $\omega_{nn'\ks}=\varepsilon_{n\ks}-\varepsilon_{n'\ks}$ and $(h_{\alpha})_{nn'\ks}$ is the matrix representation of the field-free velocity operator. For infinite-dimensional Hilbert space of Bloch states, the frequency-independent coefficient  $h_{\alpha \beta}=\partial_{\beta} h_{\alpha} = \delta_{\alpha \beta}$ is constant, independent of $\ks$. However in practical calculations, when a finite-dimensional subspace of Bloch states is used, the coefficient $h_{\alpha \beta}(\ks)$ acquires a non-trivial dependence on the quasi-momentum $\ks$. It compensates for the diminished oscillator strength and ascertains that in the limit $\omega \rightarrow 0$, the optical conductivity of the band insulator is regular and vanishes as it physically should. In the static limit, Eq.(\ref{sigma}) reduces to 
\begin{equation}
\sigma_{\beta \alpha} (\omega \rightarrow 0^+) \approx 
\frac{i}{\omega} \sum_n \int_{{\rm BZ}}  \frac{d^3 \ks}{4 \pi^3} 
\partial_{\alpha} (h_{\beta} \rho_0)_{nn\ks} =
\frac{i}{\omega} \sum_n \int_{{\rm BZ}}  \frac{d^3 \ks}{4 \pi^3} 
f_{n \ks} \left( \frac{1}{ m^{\ast}_{n \ks}} \right)_{\alpha \beta} \equiv 0
\end{equation}
where $[1/m^{\ast}_{n \ks}]_{\alpha \beta}$ is the inverse effective mass tensor  of band $n$ with crystal momentum $\ks$, and thus the apparent zero-frequency divergence in Eq.(\ref{sigma}) is removed on account of the effective mass sum rule \cite{Apostolova2020,Sipe1993,Aversa1995}. The frequency-dependent dielectric tensor is related to the optical conductivity by 
\begin{equation}
 \eps_{\alpha \beta}(\omega) = \delta_{\alpha \beta} + 4 \pi i \sigma_{\alpha \beta}(\omega) / \omega 
\end{equation}
For ZnO, $\eps={{\rm diag}}(\eps_{xx},\eps_{xx},\eps_{zz})$ is diagonal and has only two independent components (the $z$-axis points along the optical axis of the crystal). The dispersion of the dielectric function for energies below the band gap energy is shown in Fig.\ref{fig:fig2}(a-b). In Fig.\ref{fig:fig2}(a), the optical conductivity was calculated by sampling the full Brillouin zone with homogeneous $20 \times 20 \times 20$ Monkhorst-Pack mesh, and by including virtual transitions from the valence bands into and out of the two lowest conduction bands. The small difference between the two screening parameters $\eps_{xx}$ and $\eps_{zz}$ is related to the optical birefringence of ZnO. By including virtual transitions into and out of the lowest eight conduction bands, the static dielectric constant ($\omega \rightarrow 0$) reproduces quantitatively the experimental result quoted in \cite{Phillips1968}.

\subsection{Electron-hole pair excitation}  

\begin{figure}
\begin{center}
\includegraphics[width=.5\textwidth]{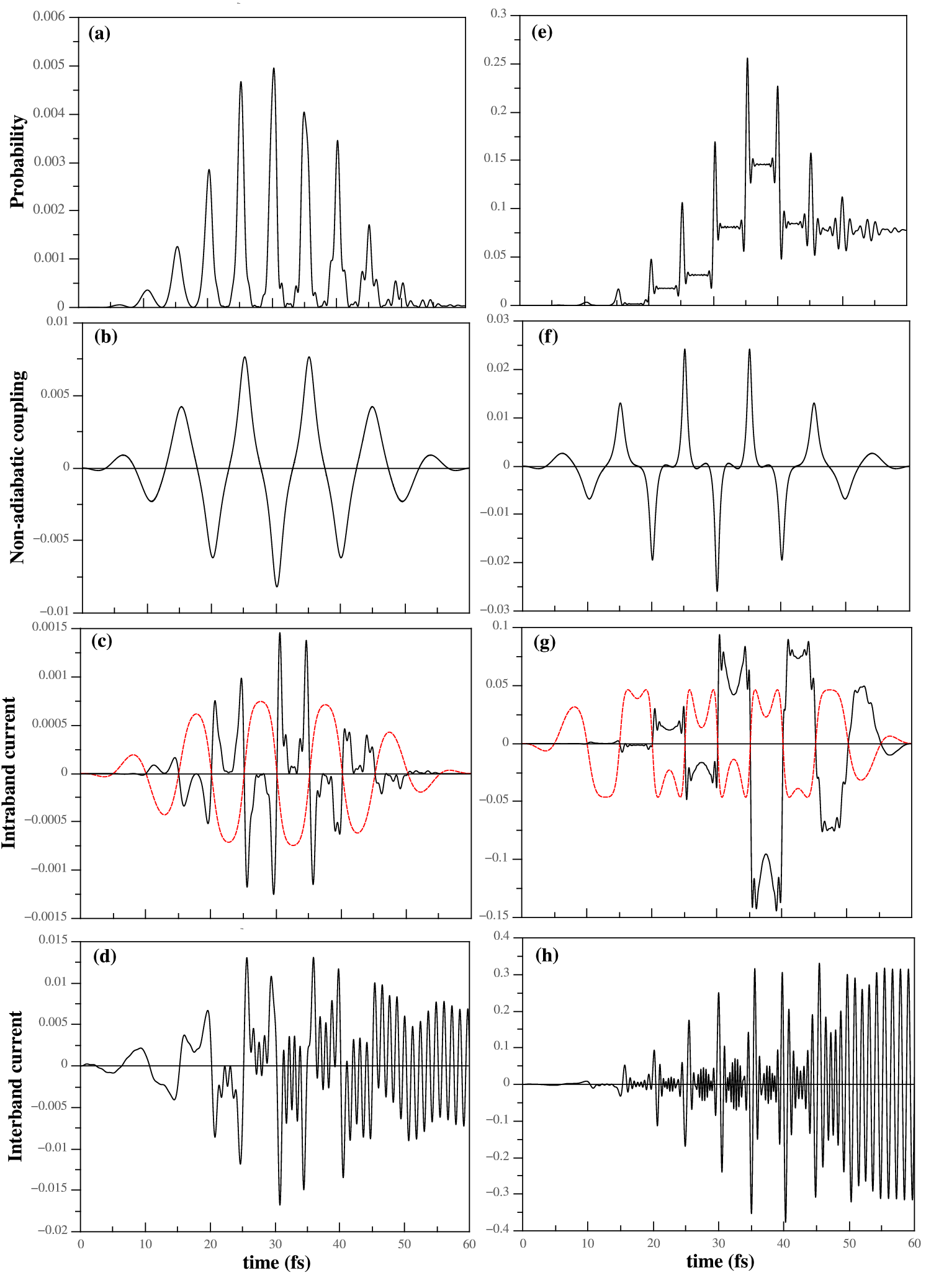}
\caption{Time-dependent response of electrons in $\Gamma$-bands 
of ZnO subjected to intense 6-cycle MIR laser pulse linearly polarized along the optical axis of the crystal, and the laser wavelength is 3 $\mu$m. In Figs.(a-d), the peak laser intensity is $10^{12}$ W/cm$^2$, and in Figs.(e-h) it is $10^{13}$ W/cm$^2$. Figs.(a,e) present the time-dependent probability for excitation of an electron-hole pair. Figs.(b,f) display the (real part of the) non-adiabatic coupling associated with the inter-band transition.  Figs.(c,g) present the time-dependent intra-band current of electrons and holes, and the dashed line traces the group velocity of an electron in the conduction band. The corresponding time-dependent inter-band currents are shown in Figs.(d,h) }   
\label{fig:fig3}
\end{center}
\end{figure}

Detailed calculations of the time-dependent self-consistent  electric field inside the bulk were made for peak laser intensities in the range $10^{11}-10^{14}$ W/cm$^2$. For laser intensities below threshold for optical breakdown $\le 10^{13}$ W/cm$^2$, the main feature in the collective response of charge carriers is the dielectric screening of the laser electric field inside the bulk. Relevant details for this type of calculation in bulk silicon can be found in Ref.\cite{Apostolova2020}. Thus we focus on single-particle aspects of electron dynamics and discuss results on photo-excitation of electron-hole pairs in ZnO for laser linearly polarized along the optical axis using the screened 
electric field $E_z(t)= (E_z)_{{\rm ext}}(t)/\eps$, here $\eps=\eps_{zz} =3.1$ is the static dielectric constant of ZnO in a model including only the two lowest in energy conduction bands, cf. Fig.\ref{fig:fig2}(a). Figs.\ref{fig:fig3}(a-h) display the temporal response of electrons in the $\Gamma$-bands with initial crystal momentum $\ks=(0,0,0)$. In this case there is one relevant inter-band coupling associated with the crystal-field split-off band  derived from the $p_z$  orbitals of O interacting with the $s$-like band of Zn orbitals. The figures present the probability for photo-ionization, the intra- and inter-band currents evaluated in the adiabatic basis of instantaneous Bloch eigen-states $|n\ks(t)\rangle$ with $\ks(t)=(0,0,A(t))$; these are given by  $P(t)=|\langle c \ks(t) | u_v({\bf 0},t) \rangle|^2$, $J^{ {\rm intra}}_z(t) = - [(h_z)_{cc \ks(t)}-(h_z)_{vv \ks(t)}]  P(t)$ and $J^{ {\rm inter}}_z(t) = - \langle c \ks(t) | u_v({\bf 0},t) \rangle 
\langle u_v({\bf 0},t) | v \ks(t) \rangle (h_z)_{vc \ks(t)} + c.c$, respectively.  In Figs. \ref{fig:fig3}(a-d), the peak laser intensity inside vacuum is $I=10^{12}$ W/c,$^2$ and the Keldysh parameter is $\gamma= \sqrt{m^{\ast} \Delta}  \omega_L/(F/\eps) \approx 1$; the reduced mass of an electron-hole pair at the Brillouin zone center is $m^{\ast}=0.08$.  In Figs.\ref{fig:fig3}(e-h) the peak laser intensity is $I=10^{13}$ W/cm$^2$, which corresponds to tunnel ionization with $\gamma=0.27$.  

For the lower intensity shown, the transient response of electrons is characterized by series of  temporally confined ionization bursts that occur near the peaks of the oscillating laser electric field, cf. Fig.\ref{fig:fig3}(a).  The temporal profile of the  non-adiabatic coupling $i \langle \dot{c}(t) | v(t) \rangle = E_z(t) (d_z)_{cv}(\ks(t))$  that drives  transitions between the valence and the conduction band is also shown in Fig.\ref{fig:fig3}(b), and $(d_z)_{cv}=h_{cv}/i \omega_{cv}$ is the transition dipole moment. Each burst leads to emergence of a small fraction of electrons in the conduction band. However a steady state is not reached, the electron distribution does not have enough time to fully equilibrate, as next ionization burst occurs. This is easily understood in the framework of the Landau-Zener model, which applies if transitions take place in a narrow time interval around the peaks of the driving laser electric field. After a pulse peak, the probability of a single transition in adiabatic basis can be written as \cite{Vitanov1996} 
\begin{equation}
 P(t > t_{{\rm peak}} ) \approx p + (1-2p) \frac{w^2}{16 (\tau^2 + w^2)^3} +
 \sqrt{p(1-p)} \frac{w}{2 (w^2+\tau^2)^{3/2}} \sin \Phi(\tau)  
\end{equation}
where $\tau = \beta (t-t_{{\rm peak}})$, the  parameters are 
$\beta^2= \sqrt{\Delta/m^{\ast}} F/2\eps$, $w=\Delta/2 \beta$ 
and $p=\exp(-\pi w^2)$ is the transition probability. The phase function is 
\begin{equation}
 \Phi(\tau)= -\frac{w^2}{2} + w^2 \ln \left(\frac{1}{\sqrt{2}} (\tau + \sqrt{\tau^2 +w^2}) \right) + \tau \sqrt{\tau^2 + w^2} + \frac{\pi}{4} + {{\rm arg}} \left[\Gamma(1-i \frac{w^2}{2})\right] 
\end{equation}
here $\Gamma(z)$ is the Euler's Gamma function, and $\Phi_S=\Phi(0)$ is the Stokes phase associated with the non-adiabatic transition.  The expression for the transition probability includes two distinct contributions: the first contribution tends to the final transition probability $P(\infty)=p$ when $\tau \rightarrow \infty$ and the second term presents transient decaying oscillations after passing the pulse peak. The transient oscillations are manifestation of the Stokes phenomenon, which can be explained by the asymptotics of the parabolic cylinder functions based on two exponentials, whose mutual interference produces them \cite{Lim1991}. The time at which the non-oscillatory part of the transition probability $P(t)$ becomes equal to $(1+\eta) p$, defines a jump time, where $\eta>0$ is a positive infinitesimal number. The oscillatory part defines a relaxation time, as the time interval over which the amplitude of oscillation is damped to a small value $\eta p$; the time duration of a single transition is $t_{{\rm jump}}+t_{{\rm relax}}$. In the adiabatic limit with $w^2 \gg 1$, the transition times are approximately given by  
\begin{equation}
 \beta t^{{\rm jump}} \approx \left(\frac{4}{\eta}\right)^{1/6} w^{1/3} e^{\pi w^2/6} ,  \quad \beta t^{{\rm relax}} \approx \left(\frac{1}{ 2 \eta} \right)^{1/3} w^{1/3} e^{\pi w^2/6} 
\end{equation}
thus the time duration of each transition increases exponentially with decrease of the field strength. In the considered example $w^2=3.56$,  $\tau^{{\rm jump}} \approx \tau^{{\rm relax}} \approx 10$ fs are nearly equal to one period of the laser oscillations $2 \pi/ \omega_L$. The transition probability does not reach its asymptotic value $\exp(-\pi w^2)$ at the end of each half-cycle, and thus the transition is not completed before the start of the next half-cycle.  

The  transient features in the transition probability 
directly translate into the intra- and inter-band currents as shown in Fig.\ref{fig:fig3}(c-d). The intra-band current is temporary modulated by the profile of the transition probability during each half-cycle. The alternation of the current is because charge carriers reverse their velocity at the times when the kinetic quasi-momentum passes through the $\Gamma$ point. The abrupt change of the electric current at the time of ionization produces a continuum bremsstrahlung-like radiation during each half-cycle, and as the number of half-cycles increases, harmonics of the fundamental laser frequency emerge from this bremsstrahlung spectrum through interference in time of emissions from all half-cycles, cf. also Ref.\cite{Protopapas1996}. During each half-cycle, the build-up of coherent superposition of states in the valence and the conduction band creates a rapidly oscillating dipole, cf. Fig.\ref{fig:fig3}(d), which in turn emits photons with energies near the bandgap energy. Like the intra-band current, the time-dependent characteristics of the inter-band current are modulated by the sharp transient features in the transition probability on a sub-femtosecond timescale,.  While the intra-band current depends linearly on the transition probability $J^{ {\rm intra}} \sim P$, the inter-band part $J^{ {\rm inter}} \sim \sqrt{ P(1-P)} \approx P^{1/2}$ when $P \ll 1$. Thus when depletion of the ground state can be neglected, the inter-band current gives the main contribution to HHG in the bulk of ZnO.

For increased laser field strength, photo-excitation of charge carriers inside the bulk becomes dominated by intense tunneling transitions which occur near the peaks of the laser electric field, when the time-dependent probability for electron-hole pair excitation changes in a step-like manner. The time duration of each transition is a small fraction of the laser half-cycle, so that each transition is completed within one half-cycle and thus successive tunneling transitions do not overlap temporary. Following each transition, a steady-state is reached, the electron-hole pairs evolve adiabatically in their respective bands: a coherent superposition of states exists in the valence and the conduction bands and only the relative phase $S(t)= \int^t dt' \Delta (\ks(t'))$ between the two states changes, here $\Delta(t)=\omega_{cv}(\ks(t))$ is the instantaneous bandgap energy. The dynamics of a single transition can be decomposed into three distinct steps: i) adiabatic time evolution from time $t_i< t_0$ prior to the pulse peak at $t_0$, ii) non-adiabatic step-like transition at the pulse peak $t_0$ followed by iii) adiabatic evolution until the time $t_f>t_0$ after the pulse peak. In this regime, the so-called adiabatic impulse model is applicable \cite{Ivakh2023}, the corresponding time-evolution operator for a single transition in the two-band model under consideration can be written as   
\begin{equation}
 U(t_f,t_i) \approx U_a(t_f,t_0) N U_a(t_0,t_i),
\end{equation}
where the adiabatic time-evolution is presented by a diagonal matrix of phase factors associated with the propagation of laser-driven holes/electrons in the valence/conduction band 
\begin{equation}
 U_a(t,t') =  \um{\exp[-i \int_{t'}^t d \tau \varepsilon_v(\ks(\tau))]}{0}{0}{\exp[-i \int_{t'}^t d \tau \varepsilon_c(\ks(\tau))]}
\end{equation}
and a non-adiabatic transition matrix defined as  
\begin{equation}
 N =  \um{\sqrt{1-p} e^{- i \Phi_S} }{-\sqrt{p} }{\sqrt{p}}{\sqrt{1-p}e^{i \Phi_S}}
\end{equation}
If the times $t_i$ and $t_f$ are chosen symmetrically with respect to the position of a pulse peak, the transition probability can be written as 
\begin{equation}
P(t_f) = p + (1-2p)P(t_i) -2 \sqrt{p(1-p)}\sqrt{P(t_i)(1-P(t_i))} \cos ( 2 S(t_i) + S_0 +\Phi_S)    
\end{equation} 
here $S_0$ is a constant relative phase associated with the initial state $|u_v({\bf 0},t_i) \rangle=\sqrt{1-P(t_i)}|v(\ks(t_i))\rangle + \sqrt{P(t_i)} e^{i S_0} |c(\ks(t_i))\rangle$ prior to the transition. Depending on the incoming phase $S_0$, the number of electrons in the conduction band may either grow or decrease after a pulse peak as shown in figures.   

During their adiabatic evolution, the charged particles are subject to the oscillating laser electric field, change speed and move with their group velocities, cf. Fig. \ref{fig:fig3}(g). The non-linear dispersion of the group velocity introduces characteristic transient features in the intra-band current: for instance at the times when the band curvature changes sign, electron-hole pairs decelerate and reduce their speeds. Thus for strong and low-frequency laser electric fields electrons and holes perform dynamical Bloch oscillations during each half-cycle of the driving pulse. While moving in their respective bands, the coherent superposition of states in the valence and  the conduction band creates a rapidly oscillating dipole, whose frequency  increases gradually with the increase of the field strength, cf. Fig.\ref{fig:fig3}(h). Near the pulse peak, when the kinetic momentum reaches an edge of the Brillouin zone,  the  oscillating electric dipole emits short-wavelength photons with energies near the cut-off energy.

\subsection{Adiabatic iteration}

To analyze this further and clearly separate the relevant time scale associated with inter-band transitions,  we iteratively construct a family of time-dependent  Hamiltonians according to \cite{Berry1987,Berry1990} 
\begin{equation}
 H_{m+1}(t)= U^{\dagger}_m(t) H_m(t) U_m(t)-i U^{\dagger}_m \dot{U}_m 
 \label{h1}
\end{equation}
where $U_m$ is an unitary operator of transformation between the static and moving Bloch bases    
\begin{equation}
 U_m(t)| n \rangle= |n_m(t) \rangle, \quad H_m(t) | n_m(t) \rangle = \varepsilon_m (n,t) | n_m(t) \rangle,
 \label{h2}
\end{equation}
here $\varepsilon_m (n)$ and $| n_m \rangle$ are the instantaneous eigen-energies and eigen-vectors of $H_m$, respectively. By demanding that the instantaneous eigen-states are parallel transported and satisfy $\langle n_m | \dot{n}_m \rangle = 0$, the matrix representation of $H_{m+1}$ in static Bloch state basis can be written as 
\begin{equation}
 \langle n'  | H_{m+1} | n \rangle = \varepsilon_m (n,t) \delta_{nn'} - i \langle n'_m(t)  | \dot{n}_m(t) \rangle (1-\delta_{nn'}), 
 \label{h3}
\end{equation}
Eqs.(\ref{h1}-\ref{h3}) define re-normalization map of Hamiltonians $H_m \rightarrow H_{m+1}$. For each $m$, we solve the time-dependent Schr\"{o}dinger equation 
\begin{equation}
 i \partial_t |u_m\rangle = H_m(t) | u_m(t) \rangle 
\end{equation}
with initial condition $|u_m(0) \rangle = | v {\bf 0} \rangle$. The transition probability is $P_m(t)=|\langle c {\bf 0} | u_m (t) \rangle|^2$. In zeroth order $m=0$, we use the ordinary adiabatic Bloch eigen-states $|n_0(t)\rangle$. Under this adiabatic   transformation, the matrix representation of the velocity 
operator $h_z$ changes according to 
\begin{equation}
 h_{m+1}(t) = U^{\dagger}_m (t) h_m(t) U_m(t) 
\end{equation}
and the total photo-current of electron-hole pairs, which is $m$-independent is given by  
\begin{equation} 
 J(t) = \sum_{n,n'} \langle u_m(t) | n \rangle [h_m(t)]_{nn'} 
\langle  n' | u_m(t) \rangle. 
\end{equation}
Note that the adiabatic eigen-states $|n_0(t)\rangle$ describe  virtual transitions between valence and conduction bands to all orders in the gauge vector potential $A(t)$, and the non-adiabatic coupling $V_0(t)=F(t) d(t)$ drives inter-band transitions between them. The description of the effective light-matter interaction can be refined in terms of new states $|n_1(t)\rangle$ with eigen-energies $\varepsilon_1(n,t)$, which incorporate virtual transitions between adiabatic states $|n_0(t)\rangle$ to all orders in the electric field strength $F(t)$. The transformation can be repeated, and new eigen-states $|n_2(t)\rangle$ are obtained, etc. None of the instantaneous eigen-states $|n_m(t)\rangle$ describe  inter-band transitions with creation of real electron-hole pairs. 

\begin{figure}
\begin{center}
\includegraphics[width=.5\textwidth]{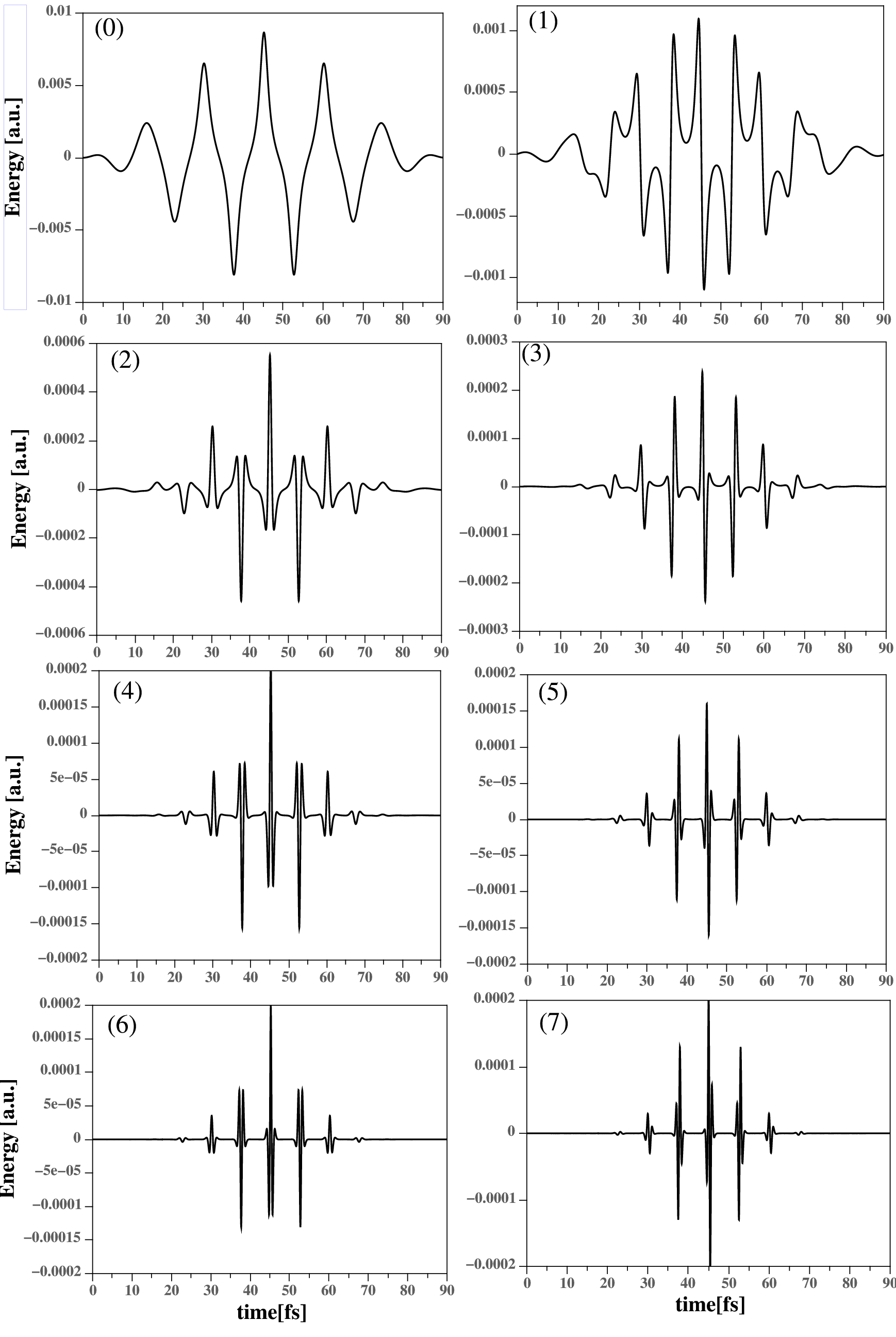}
\caption{Time-dependent non-adiabatic coupling associated with inter-band transition at the $\Gamma$-point in ZnO as a function of the adiabatic iteration number $m=0 \ldots 7$ (cf. also text). The crystal was subjected to 6-cycle MIR pulse linearly polarized along the optical axis of the crystal, the laser wavelength is 
4.5 $\mu$m laser wavelength, and peak laser intensity $10^{12}$ W/cm$^2$.}   
\label{fig:fig4}
\end{center}
\end{figure}

\begin{figure}
\begin{center}
\includegraphics[width=.5\textwidth]{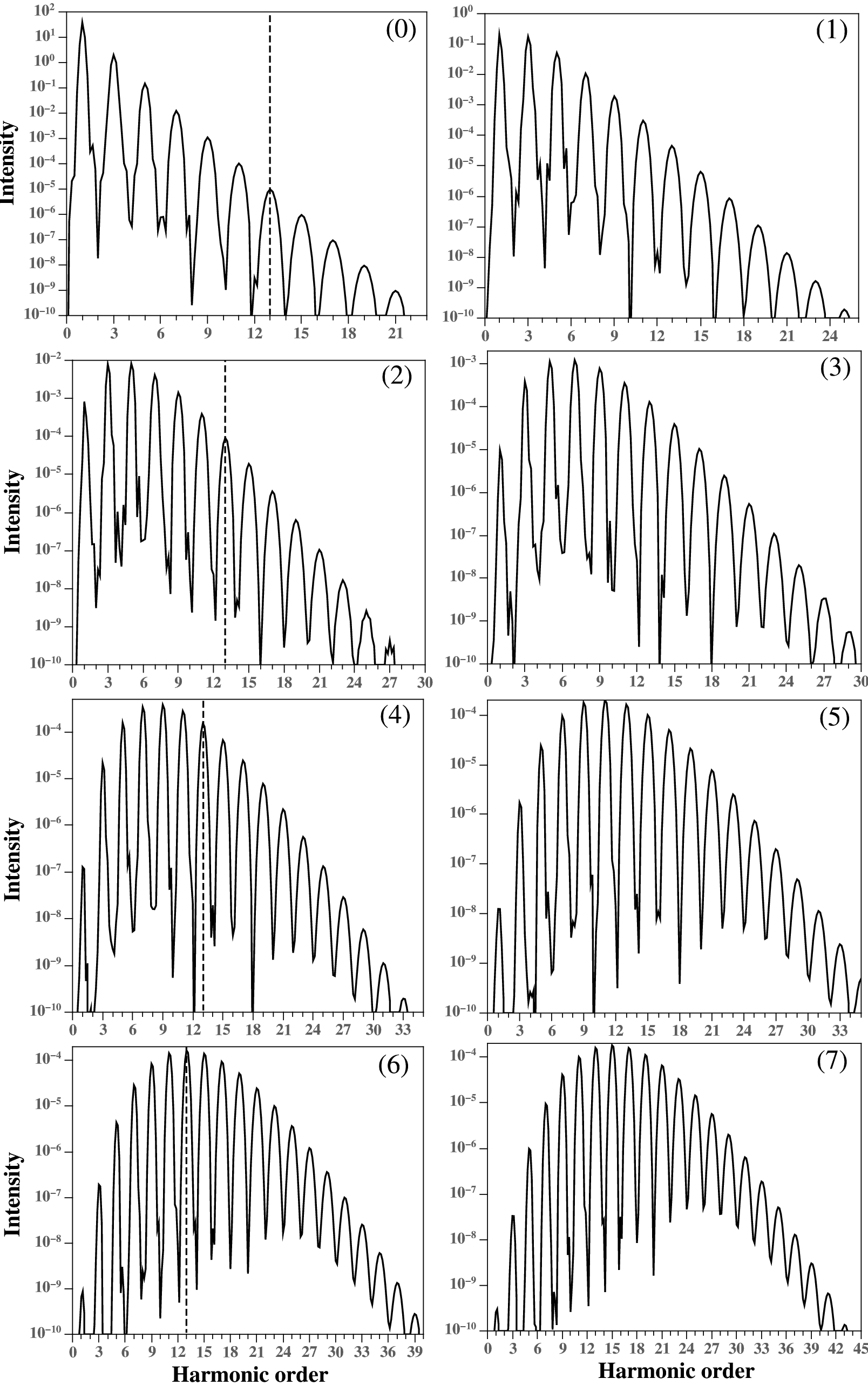}
\caption{Power spectrum of the non-adiabatic coupling in Fig.(\ref{fig:fig6}) as a function of the adiabatic iteration number $m=0 \ldots 7$ (cf. also text).}   
\label{fig:fig5}
\end{center}
\end{figure}

In Fig.\ref{fig:fig4} we present the time-dependence of the iterated non-adiabatic coupling for each $m$. During the first two iterations the time-dependent coupling decreases gradually in magnitude and develops transient sub-cycle structure. This is easily understood from the expression for the Fourier transform of the coupling   
\begin{equation}
 V_m(t) = i \langle c_m (t) | \dot{v}_m(t) \rangle    \approx \frac{1}{\Delta^m} \frac{d^m}{dt^m} V_0(t) = \int d \omega e^{-i \omega t} \left(\frac{\omega}{i \Delta} \right)^m  V_0(\omega) 
\end{equation}
with $V_0(t)=F(t) d_{cv}(t)$. The Fourier components below the band gap energy attenuate rapidly, while high-energy features with $\omega > \Delta$ are magnified.  Already in the 3-th iteration, the magnitude of the coupling $V_3 \sim 10$ meV  becomes sufficiently weak, such that the transition probability can be evaluated with high accuracy in first order of perturbation theory. By increasing $m$, the time-dependent coupling stabilizes in magnitude and displays an atto-second pulse train of bursts repeated each half cycle of the fundamental laser field. The harmonic field is composed of odd-order harmonics, harmonics below the band edge are suppressed, while the harmonics near the band edge are dominant and are of nearly equal intensity, cf. Fig.\ref{fig:fig5}(0-7). Thus an  effective Hamiltonian $H_m(t)$ is found, which describes the interaction of electrons with atto-second pulse train, and the pulse repetition frequency is 2 $\omega_L$. In the effective description of the light-matter interaction, inter-band transitions occur by resonant absorption of single highly-energetic UV photon of energy $\hbar \omega \ge \Delta$.  

\begin{figure}
\begin{center}
\includegraphics[width=.5\textwidth]{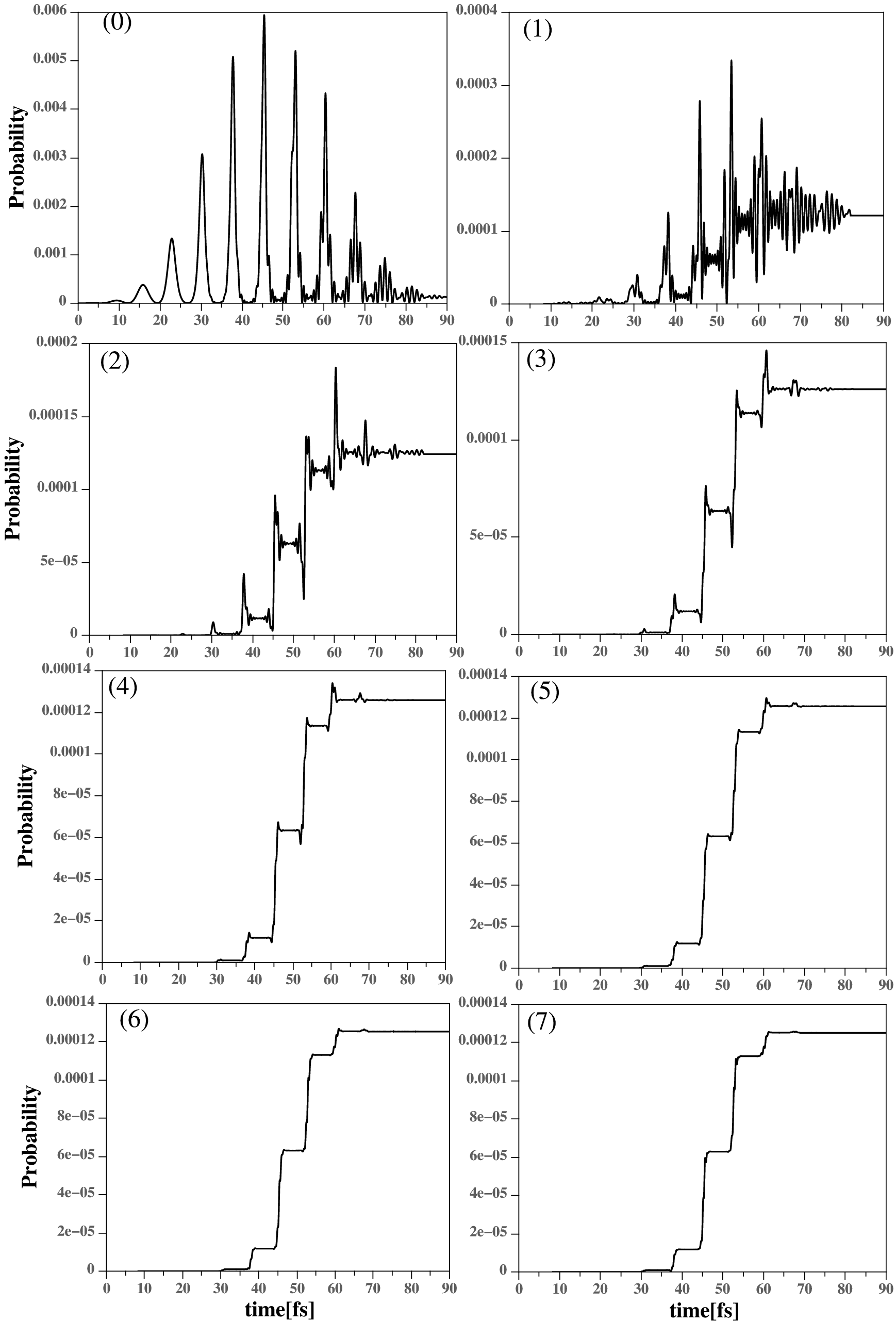}
\caption{Time-dependent response of electrons in $\Gamma$-bands 
of ZnO crystal subjected to intense 6-cycle MIR laser pulse linearly polarized along the optical axis of the crystal, the laser wavelength is 4.5 $\mu$m and the peak laser intensity is $10^{12}$ W/cm$^2$. Fig.(0-7) present different transition histories labeled by an adiabatic iteration number $m=0,\ldots,7$ 
(cf. also text)}   
\label{fig:fig6}
\end{center}
\end{figure}

The sequence of transition histories for $m=0$ to $m=7$ are shown 
in Figs.\ref{fig:fig6}. For low $m$, the transition probability displays transient oscillations associated with the Stokes phenomenon.  For $m>2$, the oscillations gradually disappear and the time-dependent probability attains universal behavior (that is independent on the detailed dependence of the Hamiltonian), which is characterized by step-wise variation near the peaks of the laser electric field. The time interval over which the transition probability changes rapidly is slightly less than 1 fs; in the considered example transitions interfere constructively, the number of electron-hole pairs grows monotonically in time by small incremental steps. 

\begin{figure}
\begin{center}
\includegraphics[width=.5\textwidth]{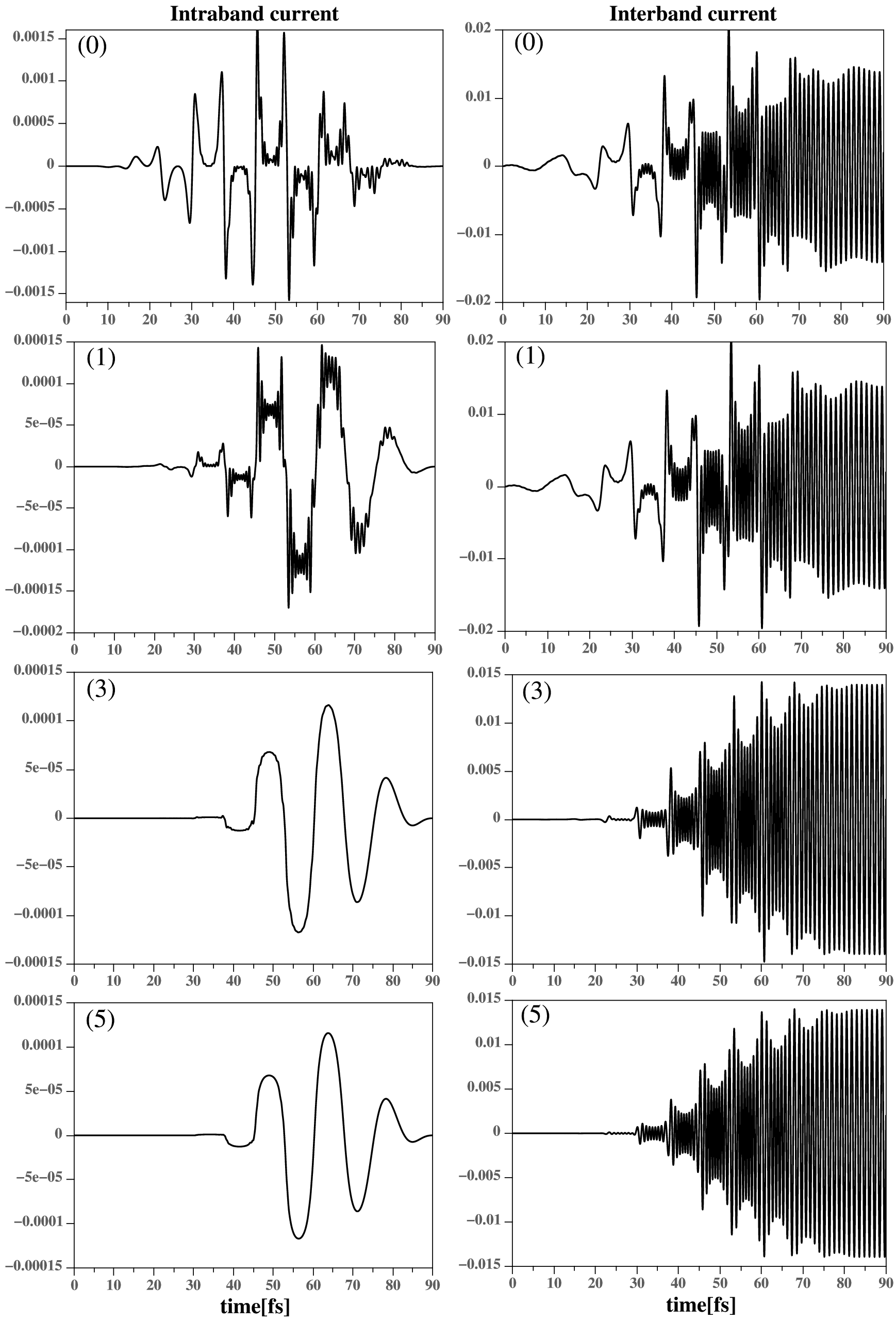}
\caption{Time-dependent intra- and inter- band currents of electrons in the $\Gamma$-bands of ZnO subjected to intense 6-cycle MIR laser pulse linearly polarized along the optical axis of the crystal, the laser wavelength is 4.5 $\mu$m and the peak laser intensity is $10^{12}$ W/cm$^2$. The label $(m)$ distinguishes different transition histories (cf. also text).}   
\label{fig:fig7}
\end{center}
\end{figure}

The transient features  associated with inter-band transitions translate into the temporal profile of the photo-currents. Figs.\ref{fig:fig7} show the time-dependent intra- and inter-band currents for $m=0,1,3$ and $5$. For $m>2$, the Brunel mechanism is clearly exposed: electrons emerge in the conduction band in short temporal bursts with zero speed around the peaks of the driving laser field, and subsequently perform laser-driven dynamical Bloch oscillations during each half-cycle. In the next half-cycle the newly photoexcited electrons join the electrons generated from previous half-cycles and the amplitude of the current gradually increases. Quite similarly, the amplitude of the inter-band current undergoes step-wise change at the times when electrons/holes emerge in the conduction/valence band. Subsequently, the super-position of states in the valence and conduction band, creates a rapidly oscillating electric dipole moment. The dipole oscillation frequency is determined by the instantaneous band-gap energy $\Delta(t)$, the oscillations are up-chirped during the first half and down-chirped during the second half of each half-cycle. 

\subsection{High order harmonic generation}

In view of the above results, the inter-band current can be approximated by  
\begin{equation}
 J(t) \approx h_{cv}(t) \sqrt{P(t)} e^{-i S(t)} + c.c. 
 \label{jt} 
\end{equation}
here $S(t)=\int^t dt' \Delta(t')$ is the quasi-classical action of an electron-hole pair, thus we neglect non-adiabatic corrections due to accumulation of Stokes phase following each transition. 
As electrons emerge in the conduction band in short temporal bursts localized near the peaks of the driving laser field, the transition probability can be approximated by a superposition of Heaviside step functions  
\begin{equation}
P(t) \approx \sum_{n=1}^{2N} (P_n-P_{n-1}) \theta(t-t_n) 
\end{equation} 
here $t_n = n \pi / \omega_L$ and $N$ is the total number of cycles. Thus the Fourier transform of the inter-band current can be decomposed into contributions from each half-cycle  according to 
\begin{equation}
 J(\omega) \approx \sum_{n=1}^{2N} \sqrt{P_n} \int_{t_{n-1}}^{t_n} dt e^{i \omega t} e^{-i S(t)} h_{cv}(t) + c.c. (-\omega)  
 \label{jom} 
\end{equation}
Since the off-diagonal matrix element of the velocity operator is a slowly varying function of time, $h_{cv}$ can be treated as constant and brought in-front of the integral in Eq.(\ref{jom}), thus we focus on the integrals 
\begin{equation}
I_n(\omega)= \int_{t_{n-1}}^{t_n} dt e^{i \omega t} e^{-i S(t)}  
\end{equation}
related to the power spectrum $|h_{cv}|^2 P_n |I_n(\omega)|^2$ of the emitted radiation inside the bulk during $n$-th half-cycle. For laser pulse of rectangular shape with identical cycles, this integral is 
\begin{equation}
I_n(\omega) = e^{i n \pi \omega/ \omega_L } 
e^{-i  n \pi \bar{\Delta}/ \omega_L}  I(\omega) 
\label{In} 
\end{equation}
where $\bar{\Delta}$ is a cycle-averaged bandgap energy 
$\bar{\Delta}=\left \langle \Delta(t) \right \rangle_{2 \pi/ \omega_L}$
and
\begin{equation}
 I(\omega) = \int_0^{\pi/\omega_L} dt e^{i \omega t} e^{-i S(t)} 
\end{equation}
gives contribution to the power spectrum from one half-cycle.
The function $I(\omega)$ is a reminiscent of the Airy function: it decays exponentially for frequencies $\omega > \omega_c$,  and it is an oscillating function for  $ \omega < \omega_c$. The cutoff energy $\hbar \omega_c$ is related to the maximal band-gap energy that be can reached by an electron-hole pair during propagation in the laser electric field. Thus for rectangular pulse shape, we obtain 
\begin{equation}
 J(\omega) \approx h_{cv} I(\omega) \sum_{n=1}^{2N} \sqrt{P_n} e^{i n \varphi} + c.c.(-\omega) 
 \label {jom2} 
\end{equation}
with $\varphi=\pi (\omega-\bar{\Delta})/\omega_L$. When  transitions interfere constructively with $P_n=n^2 p$, the partial sum in Eq.(\ref {jom2}) can be evaluated analytically giving result for the Fourier transform of the photo-current 
\begin{equation}
 J(\omega) \approx   h_{cv} I(\omega) p^{1/2}  \frac{[2 N e^{i N \varphi} \sin(\varphi/2) - \sin N \varphi]}{2i \sin^2 (\varphi/2)} e^{i N \varphi}  + c.c. (-\omega), 
 \label{comb}
\end{equation}
which shows that the interference of emissions from all half-cycles within the incident pulse duration produces a comb  of frequencies $\omega_n = \bar{\Delta} + 2 n \omega_L$,  equi-distantly spaced by $2 \omega_L$. Note that the cycle averaged bandgap energy is an odd-integer number multiple of the photon energy in case of constructive interference of transitions, which is easily understood from the result for the final transition amplitude
\begin{equation}
 a_{cv} = \int^{\infty} dt V_m(t) e^{-i S(t)} =  \sum_{n=1}^{2N} (-1)^n e^{-i n \pi\bar{\Delta}/ \omega_L}  \int_0^{\pi/\omega_L} dt V_m(t) e^{-i S(t)} = 
 \frac{\sin \left(N \pi \left[\frac{ \bar{\Delta}}{\omega_L}+1 \right]\right)}{\sin \left( \frac{\pi}{2} \left[ \frac{ \bar{\Delta}}{\omega_L}+1\right]\right)} p^{1/2}       
\end{equation}
where we have used that the non-adiabatic coupling changes sign upon time translation on one half-period of the laser oscillations. Thus the condition for constructive interference of transitions from all half-cycles is  
$\bar{\Delta}=(2n+1) \hbar \omega_L$, such that frequency comb of Eq.(\ref{comb}) is composed of odd-order harmonics of the fundamental drive laser frequency. For pulsed laser irradiation with $\sin^2$ envelope, the power spectra of emissions from each half-cycle are not identical, each emission is characterized by its own power spectrum , i.e.  $|I_n(\omega)|^2 \ne |I(\omega)|^2$ and the phase-relationship in  Eq. (\ref{In}) does not hold, which therefore results in less well defined and blurred interference pattern. The blurring of the HHG spectra for $\sin^2$ envelope is because the bandgap energy $\Delta(t) \approx \Delta [1 + \gamma^{-2} \sin^4 (\pi t /\tau) \sin^2(\omega_L t) ]^{1/2}$ is not time-periodic, while for rectangular pulse envelope in which all cycles are identical, the periodicity condition $\Delta(t) = \Delta(t+\pi/\omega_L)$ holds for the duration of the incident pulse.

\begin{figure}
\begin{center}
\includegraphics[width=.5\textwidth]{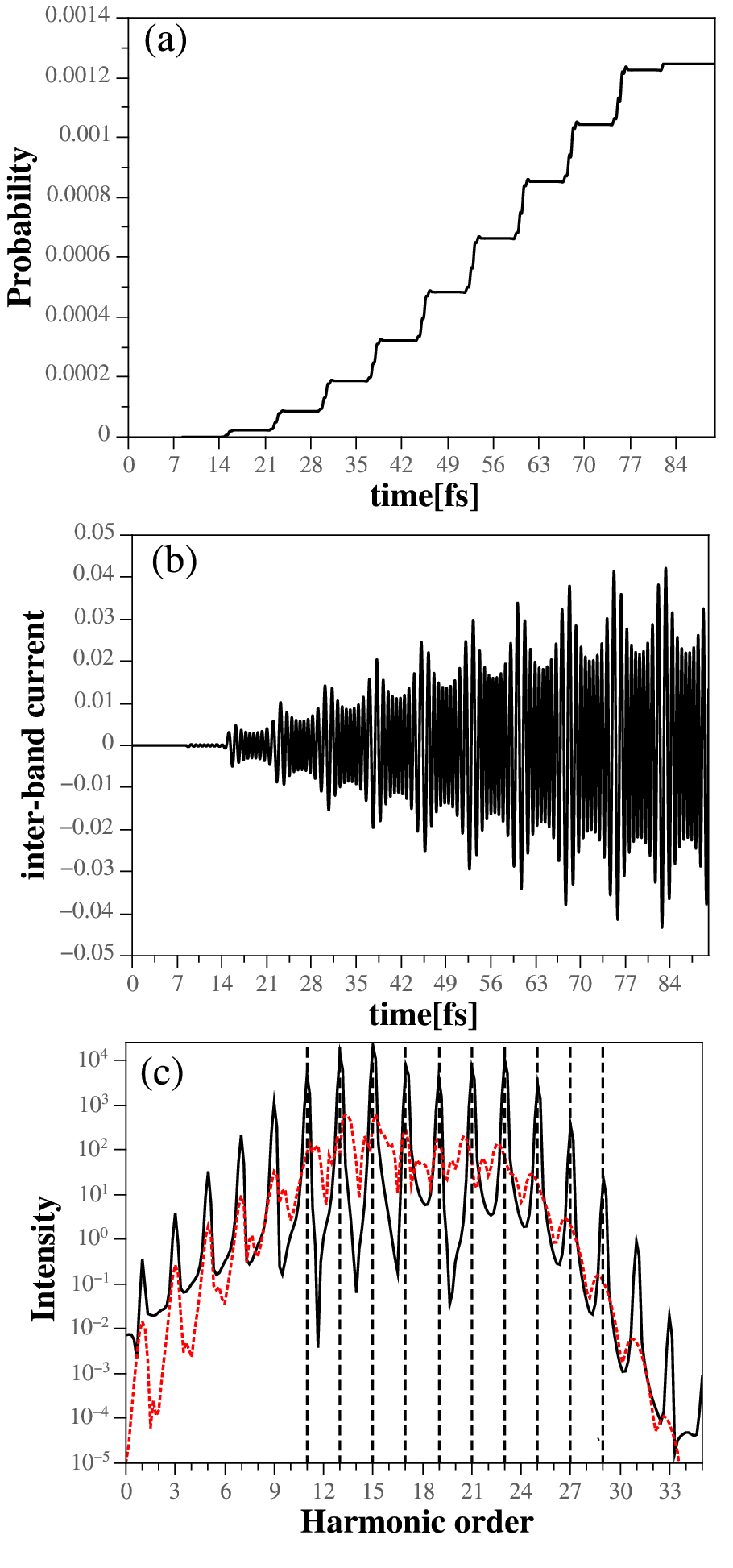}
\caption{(a-b) Time-dependent response of electrons in $\Gamma$-bands of ZnO crystal subjected to intense 6-cycle MIR laser pulse having rectangular envelope, linearly polarized along the optical axis of the crystal. The laser wavelength is 4.5 $\mu$m and the laser intensity is $10^{12}$ W/cm$^2$. Fig.(a) shows the time-dependent transition probability, Fig.(b) gives the time-dependent inter-band current and Fig.(c) presents the associated power spectrum of the inter-band current.}   
\label{fig:fig8}
\end{center}
\end{figure}

The time-dependent transition probability, the inter-band current and the associated power spectrum of the emitted radiation inside the bulk are shown in Fig.\ref{fig:fig8} for the 4.5 $\mu$m laser wavelength using a rectangular pulse envelope, the laser intensity of the driving pulse is $10^{12}$ W/$cm^2$. In this specific example, the cycle-averaged band-gap energy  $\bar{\Delta} \approx 19 \hbar \omega_L$. Each emission produces a broad continuous background with well-defined cutoff energy. Discrete high-order harmonic peaks at odd-integer numbers emerge from this background through interference in time of emissions from all half-cycles. The comparison between the power spectra for the rectangular and $\sin^2$ pulses, shows that the pulse envelope is an essential determinant for the production of HHG characterized by clean frequency comb of odd-integer harmonics of the drive laser frequency. 

\subsubsection{Dependence on the crystal momentum}

\begin{figure}
\begin{center}
\includegraphics[width=.5\textwidth]{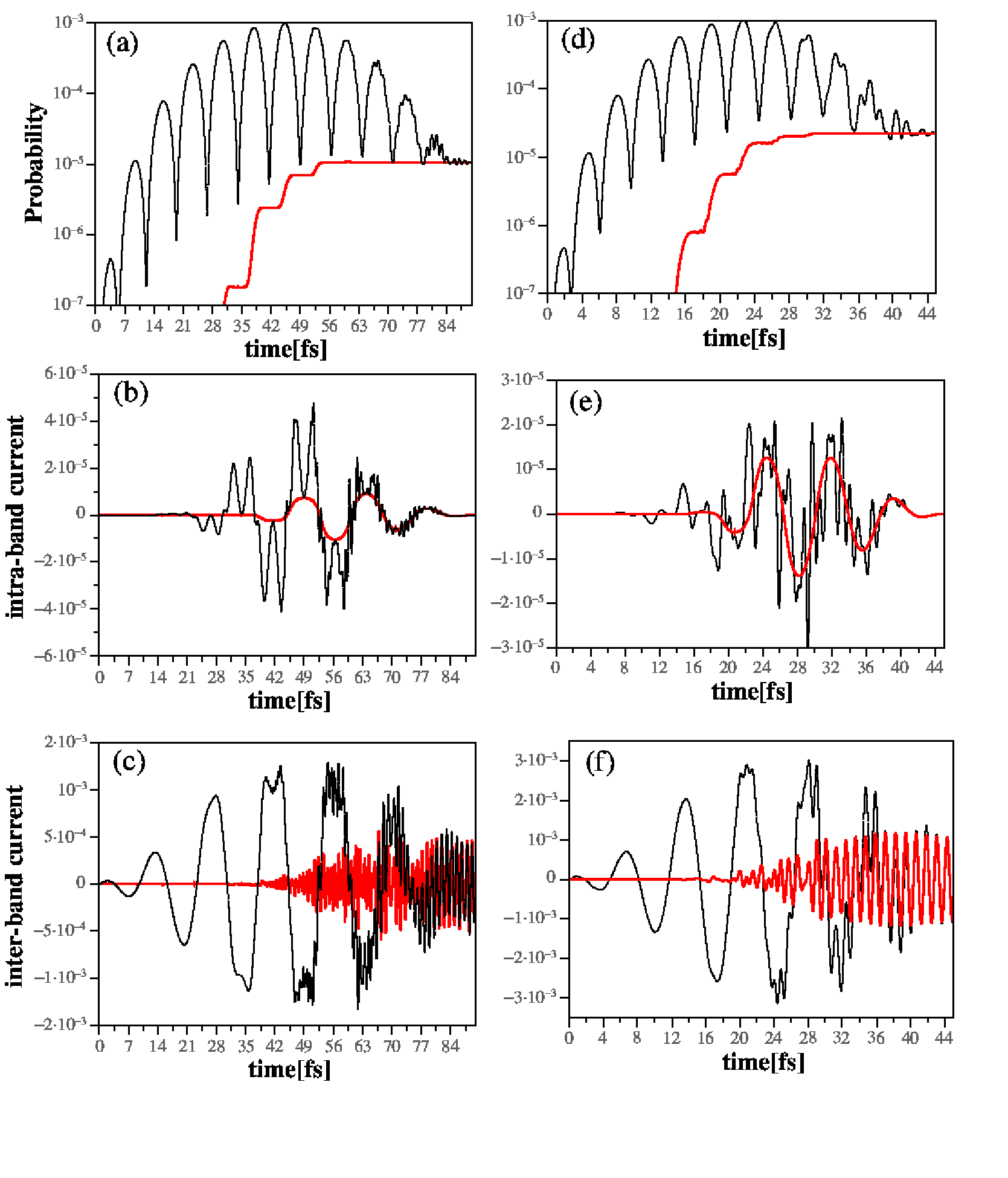}
 \caption{Time-dependent response of valence and conduction band states in ZnO subjected to intense 6-cycle laser pulse. The laser is linearly polarized parallel to the optical axis, the peak laser intensity is $10^{12}$ W/cm$^2$. The laser wave-length is 4.5$\mu$m in Figs.(a,c,e) and it is 2.25 $\mu$m in Figs.(b,d,f). The Brillouin zone was sampled by 151 points equidistantly spaced along the optical axis (the $z$-axis). The dashed lines give results in the adiabatic basis of states, and the solid lines present results in the effective description of the laser-matter interaction with $(m=6)$, (cf. text). }   
\label{fig:fig9}
\end{center}
\end{figure}

\begin{figure}
\begin{center}
\includegraphics[width=.5\textwidth]{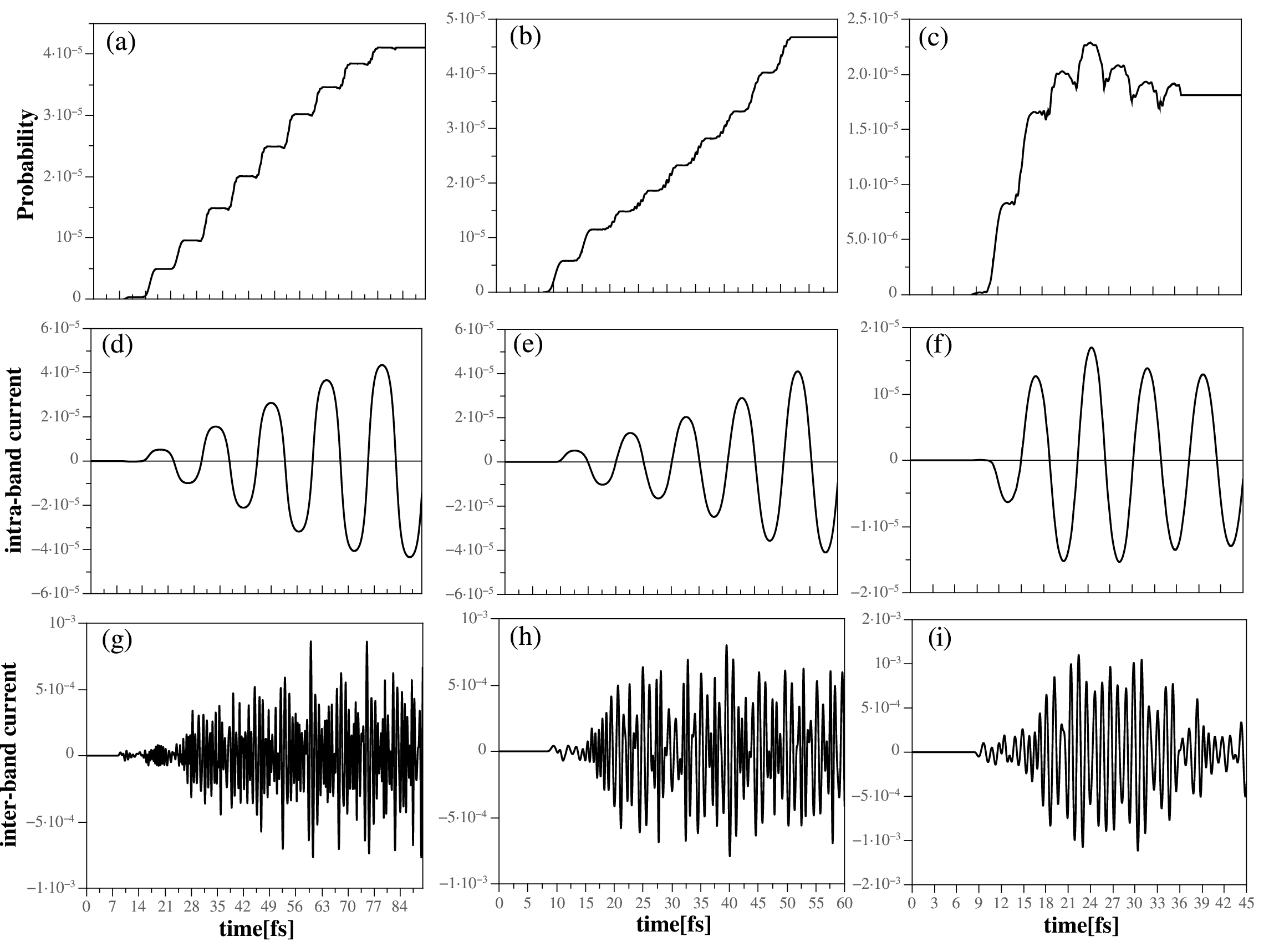}
\caption{Time-dependent response of valence and conduction band states in ZnO subjected to intense 6-cycle laser pulse with rectangular envelope. The laser is linearly polarized parallel to the optical axis, the drive laser intensity is $10^{12}$ W/cm$^2$. The laser wave-length is : 4.5$\mu$m in Figs.(a,d,g),  3 $\mu$m in Figs.(b,e,h) and 2.25 in Figs.(b,e,h). The Brillouin zone was sampled by 151 points equidistantly spaced along the optical axis (the $z$-axis). All figures present results in the effective description of the laser-matter interaction with $(m=6)$ (cf. text).}   
\label{fig:fig10}
\end{center}
\end{figure}

In previous sections we focused our discussion on electron dynamics locally in a small neighborhood of the Brillouin zone center, while the macroscopic electric current associated with HHG, involves the contributions from all $\ks$-points in the Brillouin zone. Figs.\ref{fig:fig9} and  Figs.\ref{fig:fig10}, show the time-evolution of the probability for photo-excitation, and the coherently summed ($J(t)=\sum_{n_3=1}^{N_3} J_{0,0,n_3}(t)/N_3$)) intra- and inter-band line currents generated from $N_3=151$ $\ks$-points in the Brillouin zone equidistantly spaced along the laser polarization direction. For $\sin^2$ envelope, Figs.\ref{fig:fig9}(a-f), the density of electron-hole pairs increases on the rising edge of the pulse, the main contribution to the electron yield comes from 
few transitions near the pulse peak and quickly saturates slightly afterwards. After passing the pulse peak, the inter-band current displays complicated beating pattern as a result of interference of inter-band transitions involving electron-hole pairs with different quasi-momentum. 

The time-dependent density of electron-hole pair excitations subjected to rectangular pulse  envelope, shown in Figs.\ref{fig:fig10}(a-i), increases monotonically in a step-wise manner, the magnitude of the intra-band current increases gradually with the increase of the number of half-cycles. Like the result for $\sin^2$ pulse envelope, the inter-band current exhibits complex beating pattern of different inter-band transitions, but all laser cycles contribute to the generation process. 

\begin{figure}
\begin{center}
\includegraphics[width=.6\textwidth]{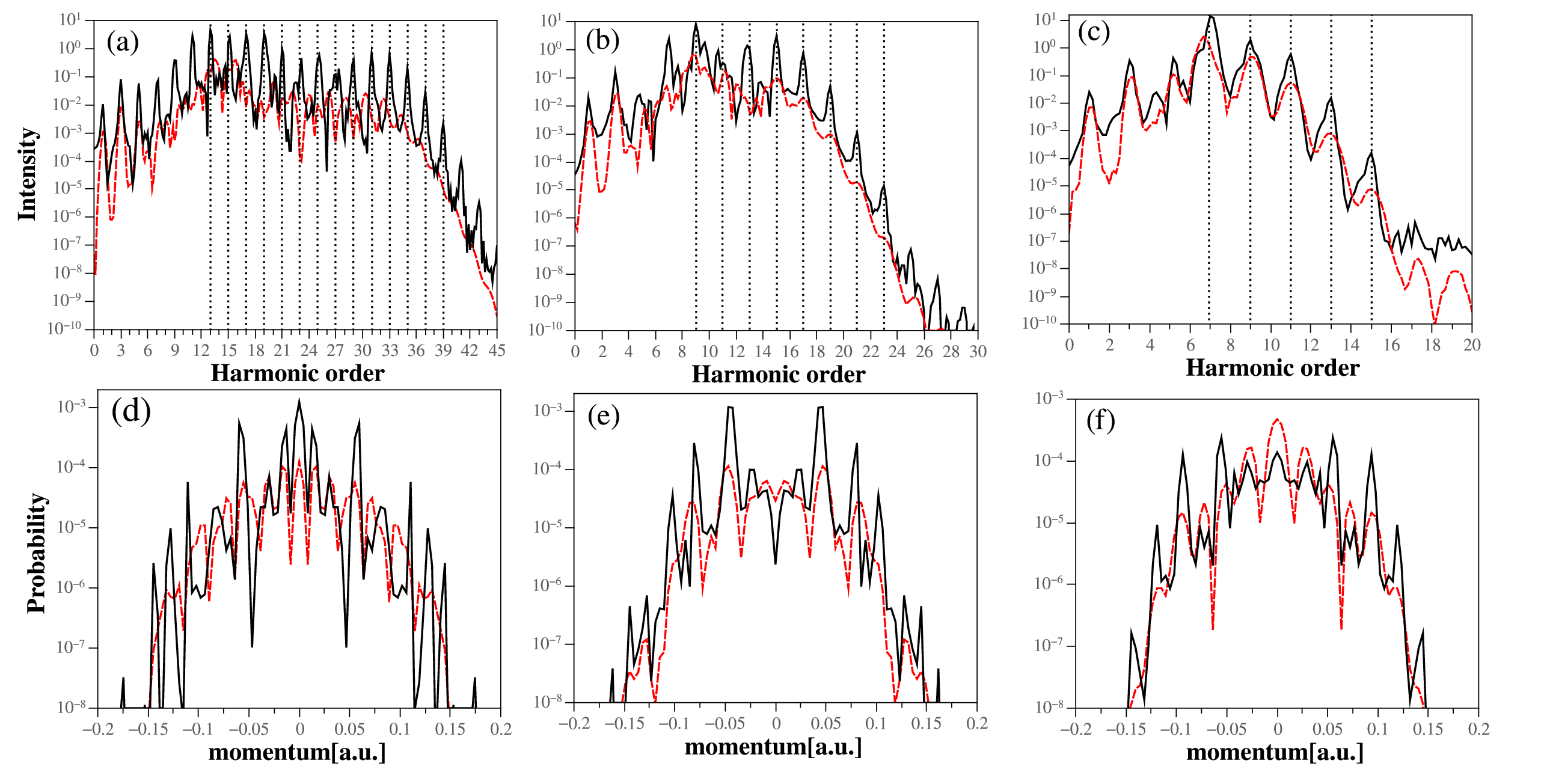}
\caption{Figs.(a,b,c) - Power spectrum of inter-band line currents in the bulk of ZnO.  Figs.(d,e,f) - final probability for photo-excitation of electron-hole pairs as a function of the initial crystal momentum of Bloch electrons. The laser wavelength is : 4.5 $\mu$m  in Figs.(a,d),  3$\mu$m in Figs.(b,e) and 2.25 $\mu$m in Figs.(c,f). The Brillouin zone was sampled by 151 points in crystal momentum, equidistantly spaced along the laser polarization direction. The ZnO crystal was subjected to intense 6-cycle MIR laser pulse the laser intensity at the pulse peak is $10^{12}$ W/cm$^2$. The solid lines correspond to to rectangular pulse envelope, while dashed lines  correspond to $\sin^2(\pi t /\tau)$ pulse envelope. }   
\label{fig:fig11}
\end{center}
\end{figure}

Figs.\ref{fig:fig11}(a-c) present the associated power spectra for both $\sin^2$ and rectangular pulse envelopes, together with the final probability for electron-hole pair excitation as function of the canonical crystal momentum $\ks$. As expected, the power spectra for the rectangular MIR pulse display clean harmonic peaks at odd-integer numbers superimposed onto the broad continuous background. The power spectra for the $\sin^2$ pulse are blurred and lack clean and well defined harmonic structure for MIR laser wavelength. For NIR laser wavelength, cf. Fig.\ref{fig:fig11}(c), the  power spectra for the two pulse envelopes look quite similarly. This is because in the regime with $\gamma > 1$,  the power spectra of emissions from nearby half-cycles are nearly identical. In this case both spectra for rectangular and $\sin^2$ pulse envelopes display well defined odd-order harmonic structure near and above the band edge, and the intensity of the high-order harmonics decreases moderately with the increase of the harmonic order. 

The final photo-electron yield is structured by sharp peaks as a result of above-threshold excitation of electron-hole pairs,  analogous to above threshold ionization (ATI) in a gas phase \cite{Agostini1979,Eberly1991}. For MIR pulse, ATI electrons emerge in the conduction band with well-defined crystal momentum. That is because of the constructive interference of inter-band transitions from different periods of the laser oscillations, i.e. when the cycle-averaged bandgap energy at a given $\ks$-point matches the energy of integer number of IR photons $\bar{\Delta}(\ks)= n \hbar \omega_L$, or equivalently when the semi-classical  action for one-cycle of the laser oscillations $S=n h$ is quantized with integer numbers multiple of the fundamental Planck's constant,  resonant-like excitation of electrons into the conduction band occurs; we shall further refer to these photo-electrons as resonant ATI electrons. 
To demonstrate that explicitly, in Figs.\ref{fig:fig12}(a-c) we show the transition histories of three different inter-band transitions at nearby points in the Brillouin zone with $\ks_1=(0,0,-0.068)$, $\ks_2=(0,0,-0.051)$ and $\ks_3=(0,0,-0.059)$. When the action quantization condition $S=20 h$ is satisfied, resonant ATI electrons with crystal momentum $\ks_3$ emerge in the conduction band with $\bar{\Delta}(\ks_3) =20 \hbar \omega_L$. The phase-sensitive probability for excitation of electron-hole pairs at nearby points with $\ks_1$ and $\ks_2$ is lower by more than an order of magnitude, which is because the  quantization condition is not satisfied and thereby results in destructive interference of  contributions from the different periods of the laser oscillations. It leads to the conclusion that the time-dependent inter-band current of resonant ATI electrons gives the dominant contribution to high order harmonic generation in the bulk of ZnO.  

Because the photo-current associated with Bloch states with  quasi-momenta $\ks$ and $-\ks$ satisfies the relationship $J(\ks,t) \approx -J(-\ks,t+\pi/\omega_L)$ connecting two adjacent half-cycles, it  implies that all even-order harmonic generated by $J(\ks,t)+J(-\ks,t)$ interfere destructively and HHG is structured by odd-order harmonic peaks. Importantly, the comparison of Fig.\ref{fig:fig11}(b) to Fig.\ref{fig:fig8}(c) clearly demonstrates that the emergence of resonant ATI electrons extends the frequency comb and the cut-off energy for HHG to  higher orders. 

\subsubsection{Laser polarization-direction dependence} 

The laser-polarization direction dependence of ATI and hence HHG 
in the bulk of ZnO is presented in two-dimensional plots of the final probability for electron-hole pair excitation as a function of the components of the canonical crystal momentum $\ks$, cf. Figs.\ref{fig:fig12}(a-d) for  $\sin^2$ pulse envelope. The electron distribution is centered in neighborhood of the $\Gamma$ point and protrudes in the direction of the laser polarization. At right angles to the laser polarization direction, the distribution is structured by equi-distantly spaced narrow fringes of resonant ATI electrons. Peculiarities of the the solid state band structure are clearly revealed in the distinct response of electrons to parallel and perpendicular laser polarization with respect to the optical axis. The sharpness of the ATI fringes is much better defined for the perpendicular laser polarization, which is a result of the anisotropy of the multiple band structure of ZnO.

\subsubsection{Comparison to experiment} 

In order to reproduce the conditions in the ZnO experiment, the laser polarization direction was set perpendicular to the optical axis, the pulse length was increased to $\tau=$ 90fs and the laser intensity was varied in the range $4 \times 10^{11}- 4 \times 10^{12}$ W/cm$^2$, which corresponds to $\gamma \ge 1$. The ellipsoidal-shaped region in the three-dimensional Brillouin zone as shown in Figs. \ref{fig:fig13}(a,c) was sampled by homogeneous $23 \times 23 \times 23$ Monkhorst-Pack mesh with $|u_1| < 1/8$, $|u_2| < 1/8$ and $|u_3| < 1/4$.  All eight valence bands together with the eight lowest in energy conduction bands were included in the expansion of the time-dependent wave-functions over static Bloch orbitals. For the considered laser intensity range, photo-excitation creates weakly ionized plasma of electron-hole pairs with final density below $10^{17}$ cm$^{-3}$, Fig.\ref{fig:fig13}. 

The time-evolution of the electron density,  the intra- and inter-band current densities in the ordinary adiabatic basis of Bloch states (with $m=0$) are shown in Figs.\ref{fig:fig14}(a-i) for three different laser intensities, $I=5 \times 10^{11}$ W/cm$^2$ in (a,d,g), $I=1.6 \times 10^{12}$ W/cm$^2$ in (b,e,h) and $I=3 \times 10^{12}$ W/cm$^2$ in (c,f,i). The photo-currents in Figs.\ref{fig:fig14}(d-h) display regular temporal profile, which reflects the transient characteristics of virtual transitions into and out of the conduction bands. The inter-band current is the dominant component of the photo-current in the considered laser intensity range, cf. Figs.\ref{fig:fig14}(g-i), non-adiabatic effects of real electron-hole pair excitation are much weaker in magnitude and become prominent in the trailing edge of the pulse. 
The corresponding HHG spectra shown in Figs.\ref{fig:fig15}(d-f) are structured by clean odd-order harmonic peaks extending beyond the band edge, in good quantitative agreement with the ZnO experiment. Similarly to the experiment, the spectral intensity of individual harmonics increases monotonously with the increase of the laser intensity. There is also a noticeable difference, there are two frequency combs of odd-order harmonics: the primary frequency comb extends to 27-th order, inter-band transitions associated with a small fraction of resonant ATI electrons 
produces a secondary frequency comb which spans the frequency range from 30-th to 45-th harmonic. By increasing the peak laser intensity, the two initially disjoint frequency combs merge. On account of this high-energy features, the cut-off energy for HHG shifts to 50th order. High order harmonics beyond 27th order were not observed in the ZnO experiment, which may be due to re-absorption of the generated UV photons inside the bulk. 

\section{Conclusion} 

In summary, the non-linear optical response of electrons and valence band holes in zinc oxide subjected to intense few-cycle femto-second mid-infrared laser pulse was investigated in a wide range of laser intensities and laser wavelengths. A connection between atto-second pulse generation, above-threshold excitation of electron-hole pairs and high-harmonic generation in bulk solids is established, that should be helpful in resolving the currently existing controversy on the mechanism of solid-state HHG. The main microscopic mechanism of solid state HHG is linked to resonant-like excitation of electron-hole pairs in short temporal bursts around the peaks of the driving laser electric field. Even when the probability of electron-hole pair excitation is small, but under a condition of constructive interference of transitions from different periods of the drive laser pulse, the occupation of the conduction band grows monotonically in time in a step-like manner. Following each transition, electron-hole pairs born in coherent superposition of states create an oscillating electric dipole moment, whose frequency gradually changes due to dispersion of the static bandgap energy. During the time interval between each two adjacent electronic transitions, the transient dipole emits short wavelength photons in a broad spectral range, and because  the emission is time-periodic, high-order harmonics emerge inside the bulk. We also show that the pulse envelope is important determinant to obtain clean and well defined odd harmonic structure. The achieved good semi-quantitative agreement with the experiment demonstrates that the commonly assumed relevance of ultra-fast de-phasing process in solid-state HHG is not mandatory. 

\section*{Appendix} 

The matrix representation of the time-dependent Hamiltonian in static Bloch state basis $|n \ks \rangle$ can be written as 
\begin{equation}
 H(\ks,\ks(t))= S(\ks,\ks(t)) H_0(\ks(t)) S^{\dagger}(\ks,\ks(t)),
 \label{hmat}
\end{equation}
where $H_0(\ks(t))$ is a diagonal matrix ${{\rm diag}} (\varepsilon_{1 \ks(t)},\varepsilon_{2 \ks(t)}, \ldots,\varepsilon_{d \ks(t)})$ of the instantaneous eigen-energies,  $d$ labels the total number of valence and conduction electron bands, and the overlap matrix $S$ of transformation between the fixed and moving bases is given by  
\begin{equation}
S_{nm}(\ks, \ks(t))=  \langle n \ks| m \ks(t) \rangle  
\end{equation}
Only for an infinite-dimensional Hilbert space, the overlap matrix is unitary. However in practice a finite-dimensional subspace of Bloch states is used in the computation, thus the unitarity of the overlap matrix is broken, such that the Hamiltonian is no longer Hermitian. To obtain an unitary approximation for the overlap matrix, to each $\ks$-point we associate an open string of closely spaced in kinetic momentum points $\{ \ks_r=\ks+\ac_r, r=1, \ldots, N_r \}$. The arc-length of the string is $\omega_B/\pi \omega_L$ (in units of $2 \pi/a$), and $\omega_B=Fa$ is the Bloch oscillation frequency.  The overlap matrix can be expressed as  a product of overlap matrices between adjacent points 
\begin{equation}
 \langle n \ks | m \ks_r \rangle = \sum_{m_1 \ldots m_{r-1}} \langle n \ks | m_1 \ks_1 \rangle \langle m_1 \ks_1 | m_2 \ks_2 \rangle \ldots \langle m_{r-1} \ks_{r-1} | m \ks_r \rangle 
\end{equation} 
Each overlap matrix $\langle m_{s-1} \ks_{s-1} | m_s \ks_s \rangle$ was subjected to the singular-value decomposition $ S=  V \Sigma W^{\dagger}$, 
where $\Sigma={{\rm diag}} (\sigma_1,\sigma_2, \ldots,\sigma_d)$ is a diagonal matrix of the singular values, and $V$ and $W$ is a pair of unitary matrices associated with the basis transformation
between two adjacent points. The singular values are real and measure the incompleteness of the computational basis of Bloch states at a given point, and typically they all deviate slightly from unity.  Thus an unitary approximation to the overlap matrix is made by $S \approx V W^{\dagger}=U$, which establishes the connection between the Bloch states at the two nearby points $\ks_s$ and $\ks_{s+1}$ on the string. Thus the full overlap matrix is approximated by a product of unitary matrices 
\begin{equation}
 S(\ks,\ks_r) \approx \prod_{s=0}^{r-1} U(\ks_s,\ks_{s+1}),   
\end{equation}
and thereby the Hermitian property of the time-dependent Hamiltonian is restored.  During this procedure, one has to carefully take into account that initially unoccupied conduction bands touch and become degenerate in energy at some points in the Brillouin zone, so that the matrix of singular values is ill-defined at these points. When a degeneracy between conduction bands occurs, the local dimension $d(\ks)$ of the computational basis set at a given canonical quasi-momentum $\ks$-point was increased $d \rightarrow d+1$ to accommodate one more band, the procedure is stopped and resumed with larger  computational basis, until the matrix of the singular values becomes regular and close to the unit matrix at all points on the string. 

At the end of this procedure, a family of Hamiltonians $H_r(\ks)=H(\ks,\ks_r),\quad r=1,\ldots,N_r\}$ is obtained. Given the set $\{H_r(\ks)\}$, the actual time-dependent  Hamiltonian $H(\ks,t)$ was obtained by cubic spline interpolation from the kinetic momentum mesh points $\ks_r$. Next we propagate the initially occupied valence electron states forward in time be applying the Cayley transformation for small equi-distant time steps $\delta=0.03$ according to 
\begin{equation}
 |u_v(\ks,t+\delta) \rangle \approx (1-i H(\ks,t) \delta/2 )^{-1} \left(1+i H(\ks,t) \delta/2 \right)|u_v(\ks,t)\rangle. 
\end{equation}
Next, the photo-induced current in Eq.(\ref{Jkt}) is evaluated according to 
\begin{equation}
 \jc(t) = - \int_{{\rm BZ}} \frac{d^3 \ks}{4 \pi^3} {{\rm tr}} \rho(\ks,t) 
 \hs(\ks,\ks(t)).  
\end{equation}
Like in Eq.(\ref{hmat}), the matrix representation of the velocity operator in the static Bloch state basis is given by 
\begin{equation}
 h_{\alpha}(\ks,\ks(t))= S(\ks,\ks(t)) h_{\alpha}(\ks(t)) S^{\dagger}(\ks,\ks(t)),
\end{equation}
with 
\begin{equation}
 (h_{\alpha})_{nn' \ks(t)} = \sum_{\gc} (\gc+\ks(t))_{\alpha} C^{\ast}_{n,\gc+\ks(t)} C_{n',\gc+\ks(t)}. 
\end{equation}
The Brillouin zone integral was approximated by a sum over 
the Monkhorst-Pack mesh points 
\begin{equation}
 \int_{{\rm BZ}} \frac{d^3 \ks}{4 \pi^3} \approx \frac{2}{V} \frac{1}{N_1 N_2 N_3} \sum_{n_1=1}^{N_1} \sum_{n_2=1}^{N_2} \sum_{n_3=1}^{N_3} 
 \label{BZint}
\end{equation}
and $V=\sqrt{3}/2 c a^2$ is the unit cell volume.

\begin{figure}
\begin{center}
\includegraphics[width=.6\textwidth]{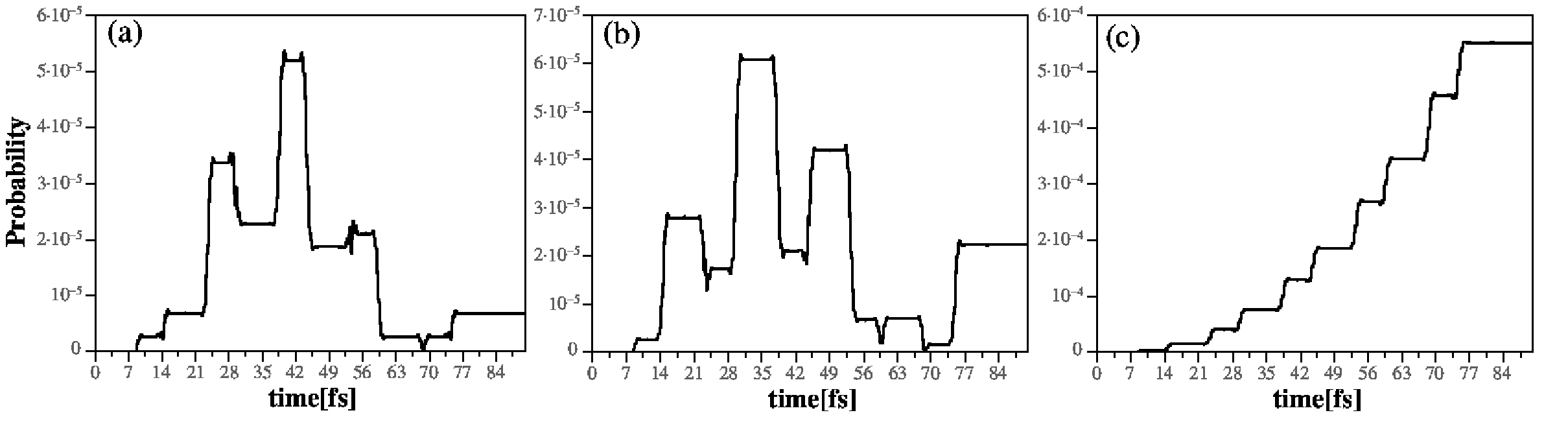}
\caption{Figs.(a,b,c) - Transition histories of  inter-band transitions at three different locations in the Brillouin zone 
$\ks_1=(0,0,-0.068)$, 
$\ks_2=(0,0,-0.051)$ and $\ks_3=(0,0,-0.59)$. The ZnO crystal is subjected to intense 6-cycle laser pulse with rectangular envelope. The laser is linearly polarized parallel to the optical axis, the drive laser intensity is $10^{12}$ W/cm$^2$ and the laser wave-length is 4.5$\mu$m.}   
\label{fig:fig12}
\end{center}
\end{figure}

\begin{figure}
\begin{center}
\includegraphics[width=.5\textwidth]{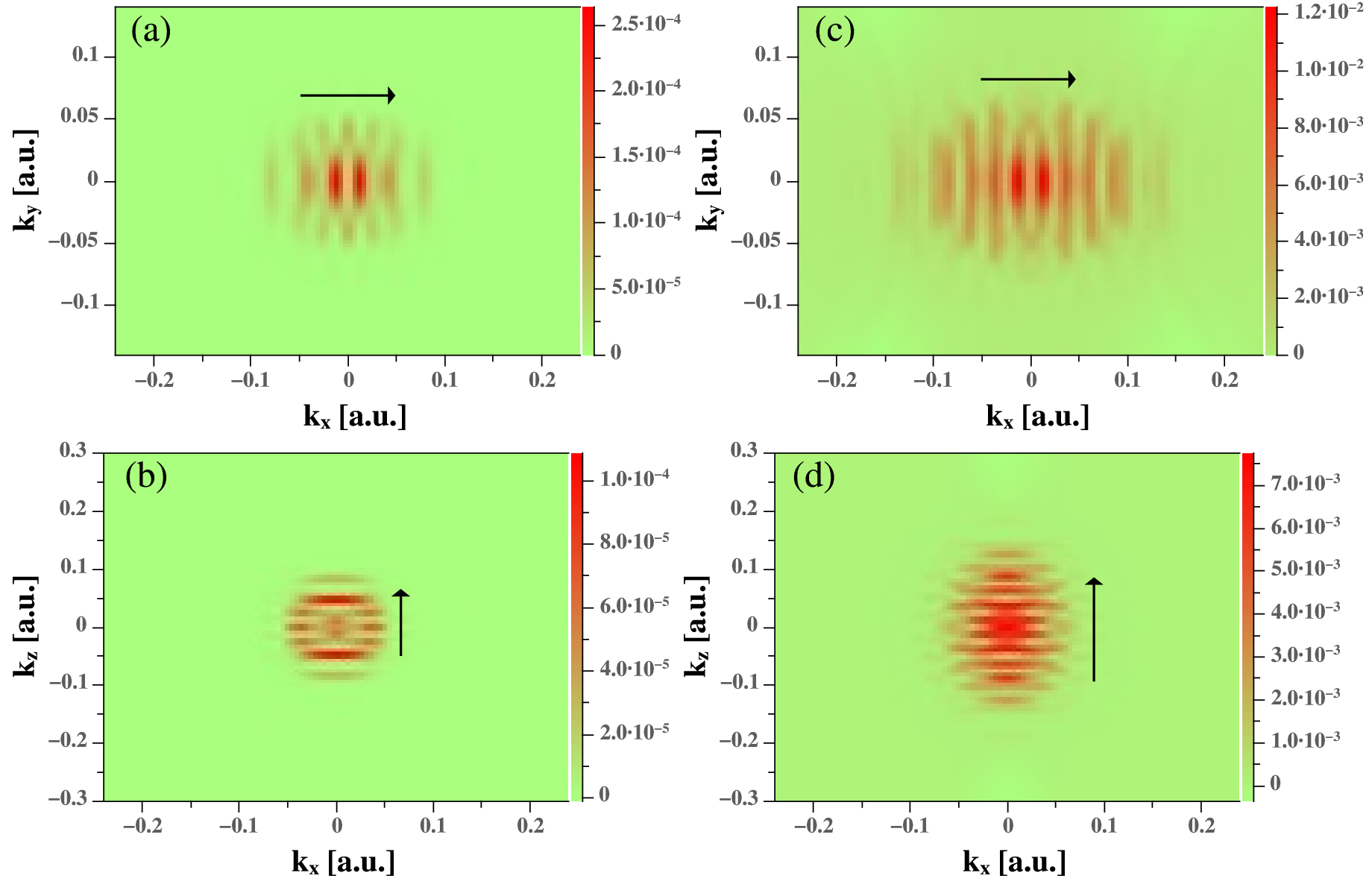}
\caption{Contour plot of the probability for photo-excitation of electron-hole pairs in ZnO as a function of the components of the crystal momentum of Bloch electrons. In Figs.(a,c) the laser is linearly polarized perpendicularly to the optical axis of the crystal. In Figs.(b,d) the laser is linearly polarized parallel to the optical axis of the crystal. The ZnO crystal was subjected to intense 6-cycle MIR laser pulse with wavelength 3 $\mu$m, the laser intensity at the pulse peak is $10^{12}$ W/cm$^2$ in Figs.(a,b), and it is  $3 \times 10^{12}$ W/cm$^2$ in Figs.(c,d). }   
\label{fig:fig12}
\end{center}
\end{figure}

\begin{figure}
\begin{center}
\includegraphics[width=.5\textwidth]{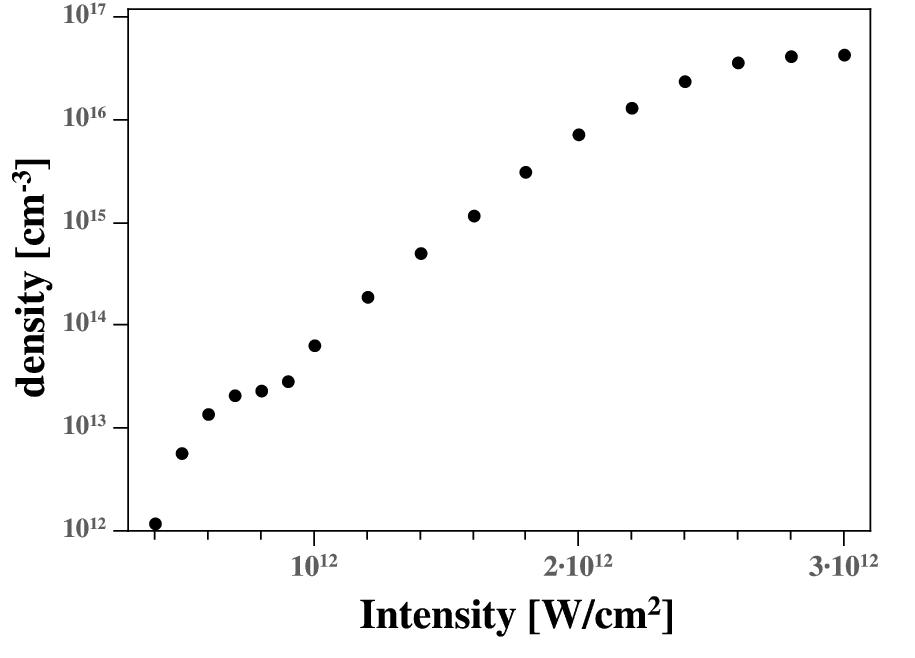}
\caption{Peak laser intensity dependence the conduction electron density (in cm$^{-3}$ ) in bulk ZnO subjected to 9-cycle pulsed laser irradiation with wavelength 3 $\mu$m. The laser is linearly polarized perpendicularly to the optical axis.}   
\label{fig:fig13}
\end{center}
\end{figure}

\begin{figure}
\begin{center}
\includegraphics[width=.5\textwidth]{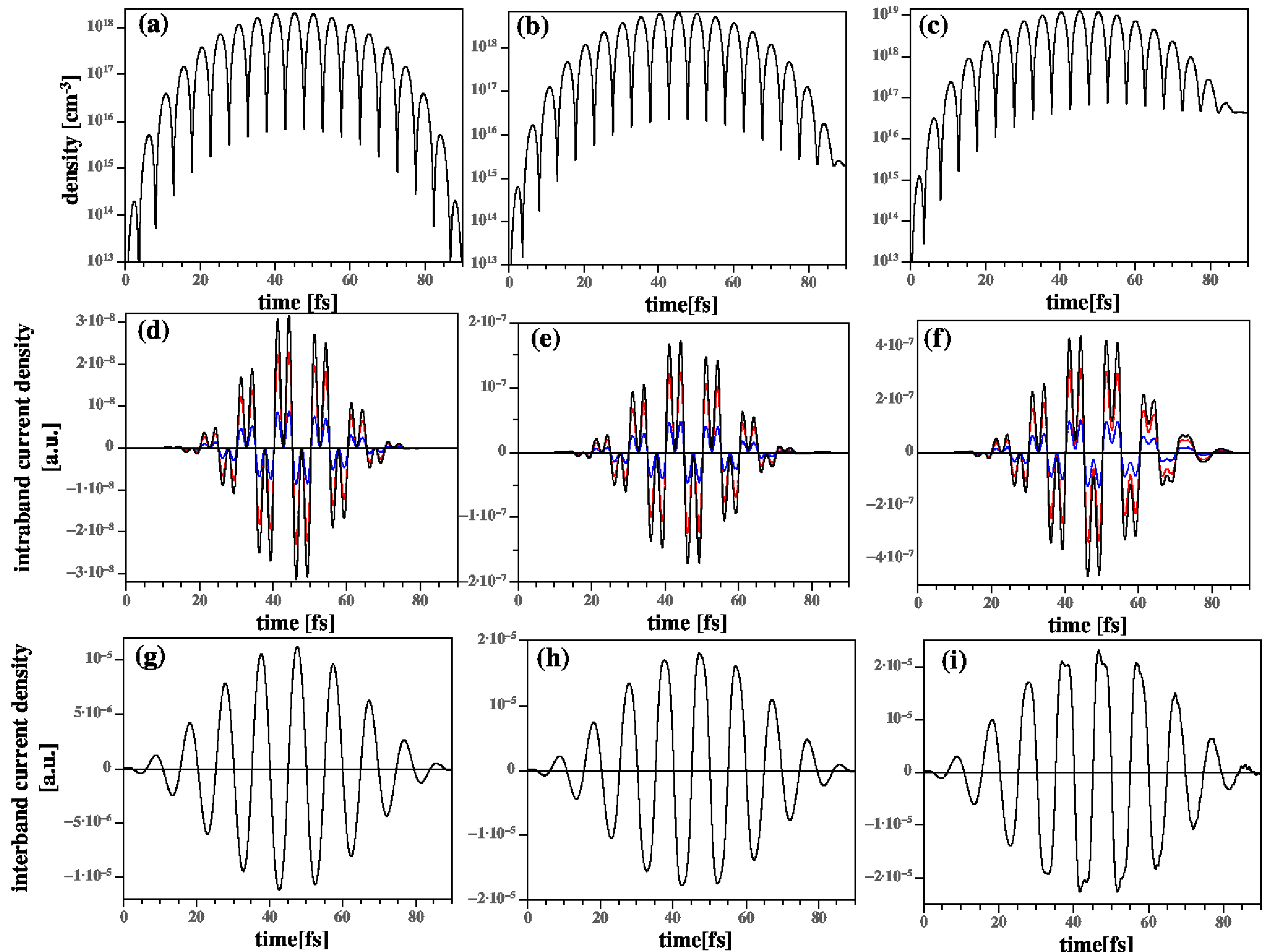}
\caption{Time-dependent response of conduction electrons and valence band holes in ZnO subjected to intense 9-cycle MIR laser pulse of laser wavelength 3 $\mu$m. The laser is linearly polarized perpendicularly to the optical axis. The peak laser intensity is  $I=5 \times 10^{11}$ W/cm$^2$ in Figs.(a,d,g),  $I=1.6 \times 10^{12}$ W/cm$^2$ in Figs.(b,e,h) and $I=3 \times 10^{12}$ W/cm$^2$ in Figs(c,f,i). }   
\label{fig:fig14}
\end{center}
\end{figure}

\begin{figure}
\begin{center}
\includegraphics[width=.7\textwidth]{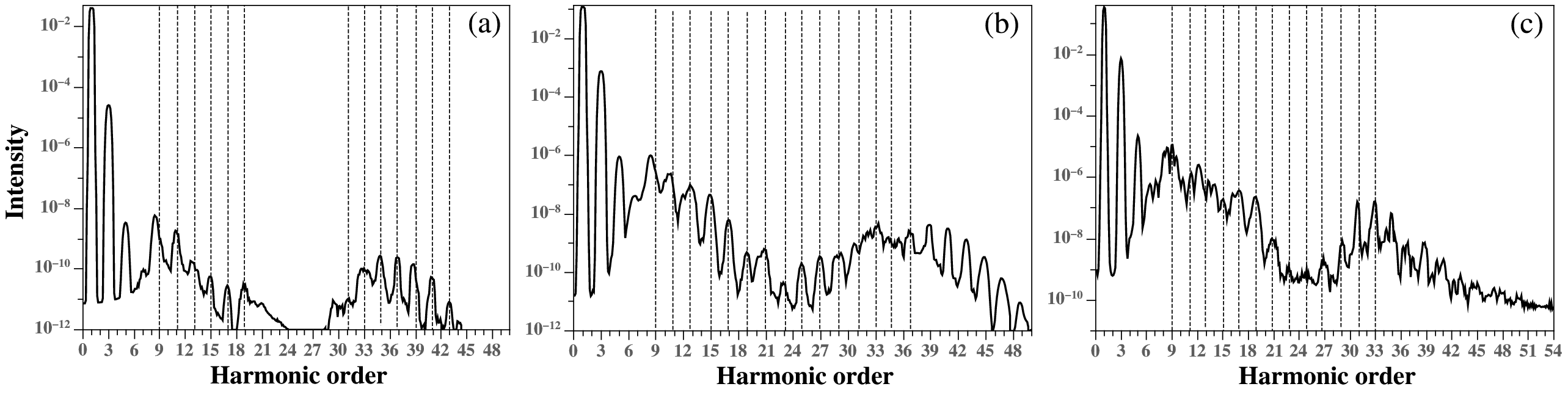}
\caption{Power spectra of high-order harmonics emitted in the bulk of ZnO. The crystal is subjected to intense 9-cycle MIR laser pulse having a wavelength 3 $\mu$m. The laser is linearly polarized perpendicularly to the optical axis. The peak laser intensity is  (a) $I=5 \times 10^{11}$ W/cm$^2$,  (b) $I=1.6 \times 10^{12}$ W/cm$^2$ and (c) $I=3 \times 10^{12}$ W/cm$^2$.  }   
\label{fig:fig15}
\end{center}
\end{figure}

\end{document}